\documentclass[letter, 10pt, conference]{ieeeconf}
\IEEEoverridecommandlockouts 
\usepackage{amsmath,amssymb,amsfonts}
\usepackage{mathtools}
\usepackage{cite}
\usepackage{algorithmic}
\usepackage{graphicx}
\usepackage[T1]{fontenc}
\usepackage[pdftex,bookmarks,hidelinks]{hyperref}
\hypersetup{linkcolor=red, citecolor=blue, urlcolor=blue}
\usepackage{textcomp}
\usepackage{bbold}
\usepackage{makecell}
\usepackage{float}
\usepackage{booktabs}
\usepackage{multirow}
\usepackage{url}
\usepackage{subcaption}
\usepackage{siunitx}
\usepackage{xcolor}
\usepackage{booktabs}
\DeclareMathAlphabet{\mathpzc}{OT1}{pzc}{m}{it}

\usepackage{physics}

\renewcommand{\mathbf}{\boldsymbol}

\newcommand{\mb}{\mathbf}

\makeatletter
\newcommand*{\rom}[1]{\expandafter\@slowromancap\romannumeral #1@}
\makeatother

\newcount\Comments  
\definecolor{darkgreen}{rgb}{0,0.5,0}
\definecolor{purple}{rgb}{1,0,1}
\definecolor{darkblue}{rgb}{0.6,0.4,0.8}

\newcommand{\kibitz}[2]{\ifnum\Comments=0\textcolor{#1}{#2}\fi}

\newcommand{\edit}[1]{\kibitz{black}      {#1}}
\newcommand{\editnew}[1]{\kibitz{black}      {#1}}
\newcommand{\etal}{,~\textit{et al.}}

\newcommand{\inv}{^{\raisebox{.2ex}{$\scriptscriptstyle-1$}}}

\begin{document}
\title{Detecting stealthy cyberattacks on adaptive cruise control vehicles: A machine learning approach}

\author{Tianyi Li$^{1}$, Mingfeng Shang$^{1}$, Shian Wang$^{2}$, Raphael Stern$^{1}$
\thanks{This material is based upon work supported by the University of Minnesota Center for Transportation Studies.}
\thanks{$^{1}$T. Li, M. Shang, and R. Stern are with the Department of Civil, Environmental, and Geo- Engineering, University of Minnesota, Minneapolis, MN 55455, USA}
\thanks{$^{2}$S. Wang is with the Department of Electrical and Computer Engineering, The University of Texas at El Paso, El Paso, TX 79968, USA}
}

\maketitle

\begin{abstract}
With the advent of vehicles equipped with advanced driver-assistance systems, such as adaptive cruise control (ACC) and other automated driving features, the potential for cyberattacks on these automated vehicles (AVs) has emerged. While overt attacks that force vehicles to collide may be easily identified, more insidious attacks, which only slightly alter driving behavior, can result in network-wide increases in congestion, fuel consumption, and even crash risk without being easily detected. To address the detection of such attacks, we first present a traffic model framework for three types of potential cyberattacks: malicious manipulation of vehicle control commands, false data injection attacks on sensor measurements, and denial-of-service (DoS) attacks. We then investigate the impacts of these attacks at both the individual vehicle (micro) and traffic flow (macro) levels. A novel generative adversarial network (GAN)-based anomaly detection model is proposed for real-time identification of such attacks using vehicle trajectory data. We provide numerical evidence {to demonstrate} the efficacy of our machine learning approach in detecting cyberattacks on ACC-equipped vehicles. {The proposed method is compared against some recently proposed neural network models and observed to have higher accuracy in identifying anomalous driving behaviors of ACC vehicles.}


\end{abstract}


\section{Introduction}\label{section1}



Automated vehicles (AVs) are expected to reshape the landscape of future transportation systems, promising many potential benefits such as enhanced traffic stability~\cite{wang2022optimal}, reduced energy consumption~\cite{sun2022energy}, and optimized parking space allocation~\cite{wang2021optimal}, among others. These advantages stem from the integration of advanced vehicular sensing, computing, and automation technologies. However, like other cyber-physical systems, AVs also present new opportunities for malicious actors to compromise vehicle safety and security~\cite{parkinson2017cyber}. Despite the myriad benefits arising from state-of-the-art vehicular sensing and automation, emerging AV technologies remain vulnerable to cyberattacks or falsified sensor measurements~\cite{petit2014potential}, which can significantly disrupt normal traffic flow and result in financial loss or even human casualties~\cite{khan2020cyber}. As a result, it is essential to address these security concerns in order to fully realize the potential of AVs in transforming transportation systems.

Although obvious attacks that cause substantial changes to vehicle behavior may be easy to detect and isolate, other vehicle compromises may be more challenging to identify, especially when the attacks change vehicle driving behavior in a subtle manner~\cite{li2018influence}. Malicious actors could introduce such attacks to fully or partially automated vehicles, including those with adaptive cruise control (ACC), via deceptive software updates that may go unnoticed~\cite{li2022detecting}. Nevertheless, even minor modifications to vehicle driving behavior can cause extensive disruptions to the transportation network by initiating new traffic jams, leading to delays, increased energy consumption, and higher emissions~\cite{dong2020impact,li2023exploring}.

While attacks can be launched on different types of AVs including ACC vehicles in various forms, they pose a significant threat to the safety, reliability, and efficiency of the transportation system as a whole. For instance, studies have demonstrated that even subtle attacks on vehicle acceleration can generate stop-and-go traffic waves and increase crash risks without directly causing vehicles to collide~\cite{wang2021stop}, thereby compromising AV safety~\cite{li2018influence}. Besides malicious attacks that directly alter AV control commands, on-board LiDAR sensor measurements are also susceptible to false data injection attacks, leading AVs to perform undesired maneuvers that {degrade} their performance~\cite{khan2020cyber}.

Various forms of cyberattacks on AVs, such as denial-of-service (DoS) attacks, message falsification, and spoofing attacks, have been shown to be able to cause string instabilities to vehicular platoons~\cite{wang2020modeling}. Even subtle attacks targeting a single vehicle can lead to significant disruption in traffic flow~\cite{khattak2021impact,wang2023novel}, resulting in reduced traffic capacity, increased energy consumption, and heightened risks of rear-end collisions~\cite{dong2020impact}. The interested reader is referred to~\cite{he2020towards} for a comprehensive discussion on the severity of different types of attacks and the potential mitigation strategies.

In view of the aforementioned potential threats caused by cyberattacks, it is necessary to understand the impacts of attacks not only on individual vehicles but also on the bulk traffic flow, and develop effective methods for real-time detection of attacks on AVs. Consequently, this will provide critical guidance in designing resilient and robust traffic control strategies to ensure safe and stable traffic flow. 

Previous studies have employed filtering theory to predict and estimate traffic states, wherein the occurrence of an attack is determined by the intersection of two state sets—the prediction set and the estimation set~\cite{mousavinejad2019distributed}. By combining a Kalman filter with a convolutional neural network, an anomaly detection approach has been developed to detect and identify anomalous behaviors of AVs~\cite{van2019real}. Recently, an efficient anomaly detection method utilizing Bayesian deep learning and discrete wavelet transform for smoothing sensor readings of AVs has been proposed to enhance the safety and security of AVs~\cite{eziama2020detection}. In the event of sensor attacks, using multiple sensors to measure the same physical variable could also increase the resilience of AVs, at the cost of creating redundancy~\cite{yang2021secure}. Besides malicious sensor attacks, falsified data injection attacks also pose significant risks to the safety and reliability of AVs. Efforts have been made to detect such attacks on AVs using a sandbox framework that facilitates isolation and evaluation of data exchanged among AVs~\cite{zhao2021detection}. Although the aforementioned studies have paved the way for developing effective methods of detecting anomalous driving behavior, many of these approaches tend to work well only with tabular data~\cite{zhao2019pyod,schmidl2022anomaly}. Furthermore, the impacts of malicious attacks on individual vehicles and the bulk traffic flow have not yet been explicitly examined.

Before the appearance of fully automated vehicles in the traffic stream, some commercially available vehicles already have low levels of autonomy features, such as ACC, which are AVs with automated longitudinal control. Conventional machine learning techniques have proven to be ineffective in detecting whether an ACC vehicle has been attacked~\cite{li2021classification,gunter2021compromised, li2023car} and lack robustness to noisy measurements or malicious attacks~\cite{li2022robustness}. Building upon our recent study~\cite{li2022detecting}, in this work, we study the impact of cyberattacks on individual ACC vehicles (i.e., AVs with low levels of automation) and the impact on the bulk traffic flow. These findings motivate us to develop trajectory anomaly detection techniques to identify compromised vehicles based on their driving behavior. Specifically, we employ car-following models to simulate normal mixed-autonomy traffic (i.e., a mixture of human-driven vehicles (HVs) and ACC vehicles) as well as abnormal traffic with a randomly selected number of compromised ACC vehicles. Using this synthetic traffic data, we then leverage a generative adversarial network (GAN)-based anomaly detection technique to identify atypical traffic conditions where traffic state evolution does not follow normal traffic dynamics. The main contributions of this work are summarized as follows:

\begin{itemize}
    \item We study the impacts of three types of attacks on individual ACC vehicles \edit{at a microscopic level}. This provides important insights into understanding the greater impact of attacks on traffic flow.
    
    \item We examine how the attacks considered may impact traffic flow at a macroscopic level. This motivates the development of effective detection strategies for improving the safety, security, and resilience of future transportation systems.
    
   \item We develop a machine learning based approach to identify ACC vehicles compromised by attacks. The proposed model is able to detect anomaly traffic in real time. Numerical results show that our model outperforms commonly used neural network models and can effectively detect all the three types of stealthy attacks on ACC vehicles. Furthermore, unlike most of the existing detection models only capable of handling tabular data, the proposed model is customized for traffic sequence trajectory data. To our knowledge, this is the first time a 1D convolutional neural network based GAN model is proposed and applied for anomaly detection on such vehicle trajectory data.
\end{itemize}




The remainder of this article is structured as follows. In Section~\ref{section2}, we introduce three types of {candidate} cyberattacks on ACC vehicles in the context of car-following dynamics. In Section~\ref{section5}, we develop a machine learning based approach for anomaly detection in compromised ACC vehicles. In Section~\ref{numerical}, we study the impacts of such attacks on individual vehicles, followed by an examination of their influence on traffic flow, and performance evaluation of the detection model. This article is concluded in Section~\ref{section6} with discussion on future research directions.

\section{Cyberattacks on ACC Vehicles}\label{section2}




There are various forms of cyberattacks that can be found in AVs~\cite{petit2014potential}, like ACC vehicles. In this section, we focus on three common types of attacks that pose a substantial risk to ACC vehicles: malicious attacks on vehicle control commands, false data injection attacks on sensor measurements, and denial-of-service (DoS) attacks, and provide a traffic model based formulation for such attacks. Although these attacks differ in nature, they can all lead to disruptive consequences for traffic flow~\cite{khan2020cyber}. 

To describe individual vehicle dynamics, microscopic car-following models are employed, where the acceleration of the vehicle of interest is related to its own state and that of the preceding vehicle, such as relative speed and inter-vehicle spacing. The general form of vehicle acceleration is given by:
\begin{equation}\label{eq:acceleration}
    a(t) = f(\mathbf{\theta}, s(t),v(t),\Delta v(t)),
\end{equation}
where $a$ is vehicle acceleration, $s$ is the inter-vehicle spacing between two consecutive vehicles, $v$ is the speed of the following vehicle, $\Delta v = v_l - v$ represents the relative speed between the following vehicle ($v$) and the lead (preceding) vehicle ($v_l$), and $\mb{\theta}$ is a vector of time-invariant model parameters. The variables, $a(t)$, $s(t)$, and $v(t)$, are time-variant. For brevity, the time index $t$ is omitted {wherever appropriate}. 




\subsection{Type~\rom{1}: Malicious Attacks on Vehicle Control}\label{attack_type1}

First, we examine malicious attacks that directly target vehicle control commands (referred to as Type~\rom{1} attacks in this study), which can be modeled as random disturbances affecting vehicle acceleration within car-following dynamics~\cite{mousavinejad2019distributed,wang2022planning,wang2023optimal}. These attacks have a direct impact on vehicle driving behavior. 

When the control command of an ACC vehicle is subject to malicious attacks, the resulting vehicle acceleration is compromised by a random signal, modeled by $\xi$, which represents the effect of the attacks~\cite{mousavinejad2019distributed,wang2022planning,wang2023optimal}. Consequently, compromised ACC vehicles may display abnormal driving behavior, such as extreme acceleration {or} deceleration, leading to disruptions in traffic flow. Without loss of generality, it is assumed that attacks can occur at any point in time. Hence, the {resulting} vehicle acceleration can be expressed as:
\begin{equation}\label{eq:att1}
    \begin{cases}
      a(t) = f(\mathbf{\theta}, s,v,\Delta v ), & \text{if unattacked} \\
      \tilde{a}(t) = f(\mathbf{\theta}, s,v,\Delta v ) + \xi, & \text{if attacked} \\
      \editnew{a(t), ~ \tilde{a}(t) \in \left[m, ~n\right]}
    \end{cases}
\end{equation}
where $\tilde{a}$ denotes acceleration of a vehicle being attacked by Type~\rom{1} attacks, $\xi$ (e.g., a variable taking a non-predefined form depending on the intention of the attackers) is the attack directly acting on vehicle control commands (acceleration), $m$ and $n$ are physical bounds of vehicle acceleration.

\subsection{Type~\rom{2}: False Data Injection Attacks on Sensors}\label{attack_type2}

The second type of cyberattacks commonly considered is false data injection attacks (referred to as Type~\rom{2} attack in this study), which can corrupt ACC sensor measurements~\cite{li2018influence,wang2022planning,wang2023optimal}. Since ACC vehicles rely on onboard sensors to measure the speed and spacing relative to the vehicle immediately ahead, falsified data could be introduced to corrupt these measurements. This type of attack can target either the hardware sensors or the software algorithms used for data acquisition in ACC vehicles, resulting in corrupted car-following data being utilized for control. Since Type~\rom{2} attacks are not directly launched on acceleration, ACC vehicles under attack are likely to experience less extreme acceleration or deceleration compared to the case of Type~\rom{1} attacks. However, attacked ACC vehicles could still exhibit anomalous driving behavior compared to using the correct input of relative speed ($\Delta v$) and spacing ($s$). This process can be characterized in the following abstract form:
\begin{equation}
   \mb{\tilde{A}} = \mb{A} + \mb{\Lambda}
   \label{eq:att2_1}
\end{equation}
where $\mb{A} = \left[s, \Delta v\right]^\top$ is the measurement vector of relative speed and inter-vehicle spacing of ACC vehicles, $\mb{\tilde{A}}$ is the corrupted form of $\mb{A}$ resulting from a vector of random variables $\mb{\Lambda}$. It is noted that $\mb{\Lambda}$ only affects attacked vehicles. In the context of car-following dynamics, this process is mathematically described as:
\begin{equation}\label{eq:att2_2}
    \begin{cases}
     a(t) = f(\mathbf{\theta}, s,v,\Delta v ), & \text{if unattacked} \\
     \tilde{a}(t) = f(\mathbf{\theta}, s + \lambda_1, v, \Delta v + \lambda_2), & \text{if attacked} \\
     \editnew{a(t), ~ \tilde{a}(t) \in \left[m, ~n\right]}
    \end{cases}
\end{equation}
where $\lambda_1$ and $\lambda_2$ are false data injection attacks on the measurements of spacing and relative speed, respectively. In other words, it follows from~\eqref{eq:att2_1} and~\eqref{eq:att2_2} that $\mb{\Lambda} = \left[\lambda_1, \lambda_2\right]^\top$ for any ACC vehicle attacked. \edit{As in~\eqref{eq:att1}, vehicle acceleration is bounded from below and above by $m$ and $n$, respectively.}

\subsection{Type~\rom{3}: Denial-of-Service Attacks}\label{attack_type3}

The third type of cyberattacks commonly seen in the literature is denial-of-service (DoS) attacks ({referred to as} Type~\rom{3} attacks in this study), where the adversaries attempt to shut down a network by temporarily or indefinitely flooding the communication traffic, resulting in communication delays~\cite{zeadally2012vehicular}. In the context of car-following dynamics, this type of attack could hinder vehicles from accessing sensor measurements by injecting dummy messages into the communication channel, causing delays in communication services. Consequently, the effects of Type~\rom{3} attacks can be mathematically characterized by a car-following model with delays in sensor measurements~\cite{wang2020modeling}, given by the following expression:
\begin{equation}\label{eq:att3}
   \begin{cases}
    {a}(t) = f(\mathbf{\theta}, s,v,\Delta v ), & \text{if unattacked} \\
    \tilde{a}(t) = f(\mathbf{\theta}, s(t-\omega),v,\Delta v(t-\omega)), & \text{if attacked} \\
    \editnew{a(t), ~ \tilde{a}(t) \in \left[m, ~n\right]}
    \end{cases}
\end{equation}
where $\omega > 0$ represents delays in sensor measurements due to DoS attacks.

In view of the three types of cyberattacks introduced above, Type~\rom{1} attacks are likely to alter vehicle driving behavior to a greater degree since they directly act on vehicle control commands, i.e., acceleration. While these attacks are different in nature, they all could potentially result in substantial disruption to traffic flow even with a portion of ACC vehicles being attacked. Hence, it is important to develop effective approaches for attack detection, which will pave the way for the development of robust traffic control strategies for future transportation systems in the presence of attacks. In what follows, we develop an effective method for cyberattack detection using generative adversarial networks (GAN).

\section{Detecting Cyberattacks via Generative Adversarial Networks~\edit{(GAN)}}\label{section5}



In this section, we propose a reconstruction-based real-time detection model designed to effectively identify subtle attacks on ACC vehicles introduced in the previous section. Specifically, we use a generative adversarial network (GAN)-based model to detect abnormal traffic, where the detection model is trained in an unsupervised learning manner~\cite{goodfellow2014generative}.

\begin{figure}[t!]
    \centering
    \includegraphics[scale = 0.11]
    {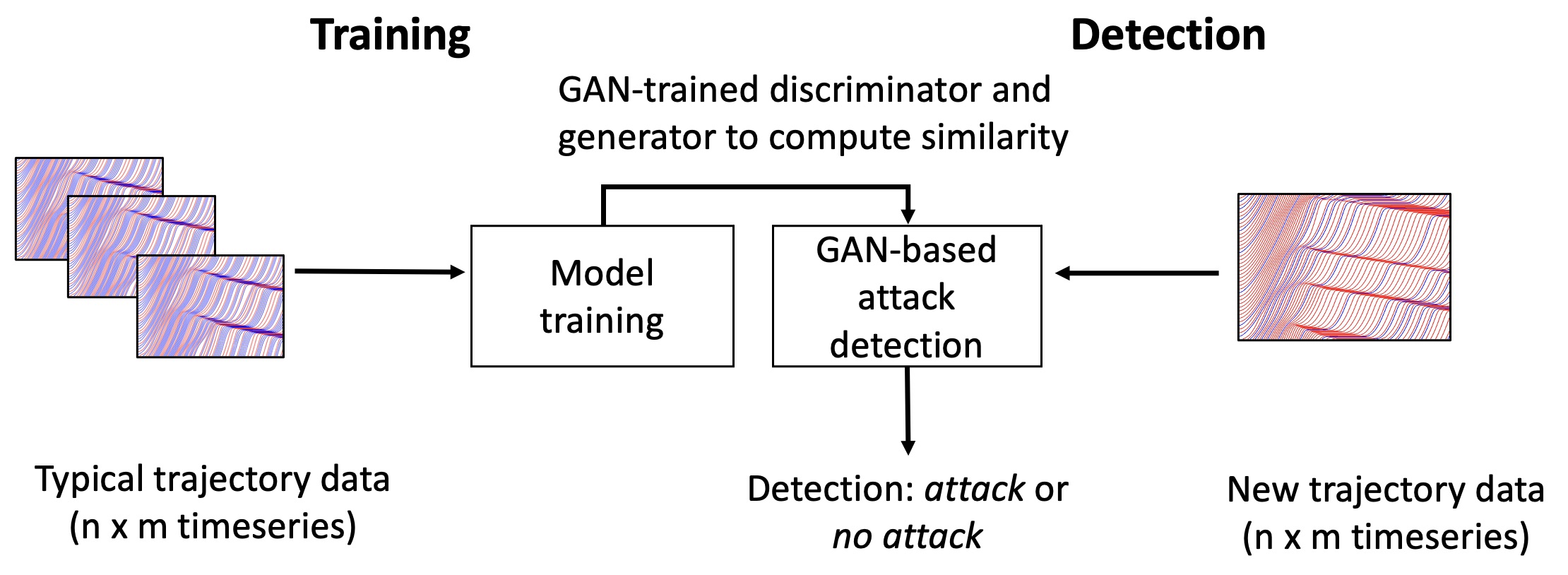}
    \caption{Overview of the proposed GAN-based attack detection strategy where typical trajectory data is used to construct a GAN for generating synthetic trajectory data, which is then applied to newly observed trajectory data to identify anomalous traffic flow.}
    \label{fig:outline}
\end{figure}

The core idea of the detection model is to first utilize the GAN to learn a representation of typical vehicle trajectories, and then test whether newly observed samples fit within the standard paradigm. Fig.~\ref{fig:outline} outlines the workflow of the detection model. Specifically, the detection model ascertains if the input trajectory data is under attack by verifying if the data conforms to the learned distribution. The proposed detection strategy includes two primary steps.

The first part involves training a GAN model capable of discovering and learning the distribution (or pattern) of normal traffic data, enabling the GAN model to generate fake yet realistic trajectory data. The second step focuses on anomaly detection, where the newly observed data is mapped back onto the latent space and reconstructed from that space. The loss, calculated by the loss function during reconstruction, is then employed to determine whether the vehicle has been attacked. In what follows, we first introduce the proposed GAN architecture in more detail. Then, we elaborate on how to leverage the GAN model for the attack detection strategy to be proposed. 
 
\subsection{Generative Adversarial Network (GAN)}

The GAN framework consists of two components: a generative model $G$ and a discriminative model $D$. The generator takes a fixed-length random variable as input to generate synthetic samples based on a learned distribution. Simultaneously, the discriminator takes an example (real or generated) as input to estimate the probability that the example originates from the domain (training data) rather than the generator. In essence, the goal of the GAN is to generate realistic samples that fit the input data distribution.

During training, both $G$ and $D$ are trained to compete against each other in an adversarial manner, playing a zero-sum game in the context of game theory. The generator tries to fool the discriminator, progressively becoming better at creating samples that appear real. Meanwhile, the discriminator continually improves its ability to identify real and generated samples. Ultimately, the process reaches an equilibrium state when the discriminator can no longer differentiate between real and generated data in every case. This means that the generator model generates plausible samples that are indistinguishable from real data.


Specifically, the generator $G$ creates synthetic samples from a latent space $\mb{Z}$, which is a projection of the input data distribution and consists of random noise vectors $\mb{z}$, such as vectors from a Gaussian distribution. Essentially, the generator's mathematical function, $G(\mb{z}, \mb{\theta}_1)$, maps the input noise vector $\mb{z}$ to the training data domain $\mb{x} \in \mb{X}$. In this study, $\mb{X}$ represents the space of normal (unattacked) traffic data, while $\mb{Z}$ denotes a latent space of a Gaussian distribution. The discriminator's function, $D(\mb{d}, \mb{\theta}_2)$, determines whether the input data originates from the training set or is generated, and outputs the probability of the input data $\mb{d}$ being real. The notations $\mb{\theta}_1$ and $\mb{\theta}_2$ represent model parameters for the generator and discriminator, respectively.
$D$ and $G$ engage in a minimax game with a value function of $V$. The loss function utilized is the cross-entropy loss expressed as $y\log(p)$, where $y$ is the true label for real data, and $p$ is the predicted probability of the observation matching the true label of $y$. The label for generated data is inverted. As a result, the $D$ improves its ability to discriminate between real and fake samples, leading to the following objective for the discriminator:
\begin{multline}\label{eq:dis}
\max_{D}  V(D)= \mathbb{E}_{\mb{x}\sim p_{\text{data}}(\mb{x})}[\log{D(\mb{x})}] + \\\mathbb{E}_{\mb{z}\sim p_{\text{z}}(\mb{z})}[1 - \log{D(G(\mb{z}))}].
\end{multline}
However, for the generator, the objective is to minimize $V(G)$ by maximizing the probability $D(G(\mb{z}))$ of being identified as real in order to deceive the discriminator. Consequently, the generator's objective function can be expressed as:
\begin{equation}\label{eq:gen}
\min_{G} V(G)= \mathbb{E}_{\mb{z}\sim p_{\text{z}}(\mb{z})}[1 - \log{D(G(\mb{z}))}].
\end{equation}


Overall, the $G$ and $D$ compete with each other in a minimax game with the value function $V(G, D)$ given by:
\begin{multline}\label{eq:gan}
\min_{G} \max_{D}  V(D,G)= \mathbb{E}_{\mb{x}\sim p_{\text{data}}(\mb{x})}[\log{D(\mb{x})}] + \\
 \mathbb{E}_{\mb{z}\sim p_{\text{z}}(\mb{z})}[1 - \log{D(G(\mb{z}))}],
\end{multline} 
where $G$ aims to minimize $V$ while $D$ tries to maximize it. The objective functions of both $G$ and $D$ are trained simultaneously. We employ stochastic gradient descent (SGD) to train the GAN model. The networks for $G$ and $D$ are trained in alternating steps until they reach a balance point where both models cannot be further improved. At this point, the generator $G$ can produce naturalistic data, and the discriminator $D$ is unable to differentiate between real and generated data.

In this study, we adopt a Convolutional Neural Network (CNN) based architecture for handling traffic trajectory data, as it is particularly suitable for time-series data modeling~\cite{fawaz2019deep}. Although the Long Short-Term Memory (LSTM) architecture is useful for capturing sequential interactions and is capable of dealing with sequence data as demonstrated in several studies~\cite{graves2013speech,li2022taxi}, CNNs have been shown to be faster, more accurate, and more computationally efficient~\cite{bai2018empirical,fawaz2019deep,wang2017time}. Therefore, we employ a 1D-CNN architecture for both the generator and discriminator models. The data involved in this study consists of multivariate time series.

In our GAN model, the generator is composed of transposed convolutional layers, batch normalization layers, and \editnew{Leaky Rectified Linear Unit (LeakyReLU)} activation layers. The generator employs transposed convolutional layers to generate a sample vehicle trajectory (acceleration, speed, and spacing) from a latent vector (input), $\mb{z}$, which is drawn from a latent space $\mb{Z}$ (a standard normal distribution in this case). The generated sample (output) has the same dimensions as the input data.
The discriminator is a CNN-based classifier, consisting of convolutional layers, batch normalization layers, and LeakyReLU activation layers. The input to the discriminator is the trajectory of vehicle speed, acceleration, and spacing information in sequence, e.g., 10 seconds of data. The discriminator uses this input to output a scalar probability that the input originates from the real data distribution. 

\subsection{Detection of Cyberattacks}
In order to detect whether an ACC vehicle has been attacked, it is necessary to map real data from $\mb{x} \in \mb{X}$ to $\mb{z} \in \mb{Z}$ in order to determine the degree of resemblance between the fake data generated by $G$ and the real ACC vehicle driving data. Our objective is to find a $\mb{z} \in \mb{Z}$ that is most similar to the given data $\mb{x}$, with the level of similarity calculated using the trained generator $G$. The generator can be considered an implicit model that reflects the distribution of normal (unattacked) traffic data, and is trained to learn a mapping $G(\mb{z}): \mb{Z} \rightarrow \mb{X}$, which maps the latent space sample $\mb{z}$ to the normal data $\mb{x}$.

\begin{figure}[t!]
    \centering
    \includegraphics[scale = 0.1]{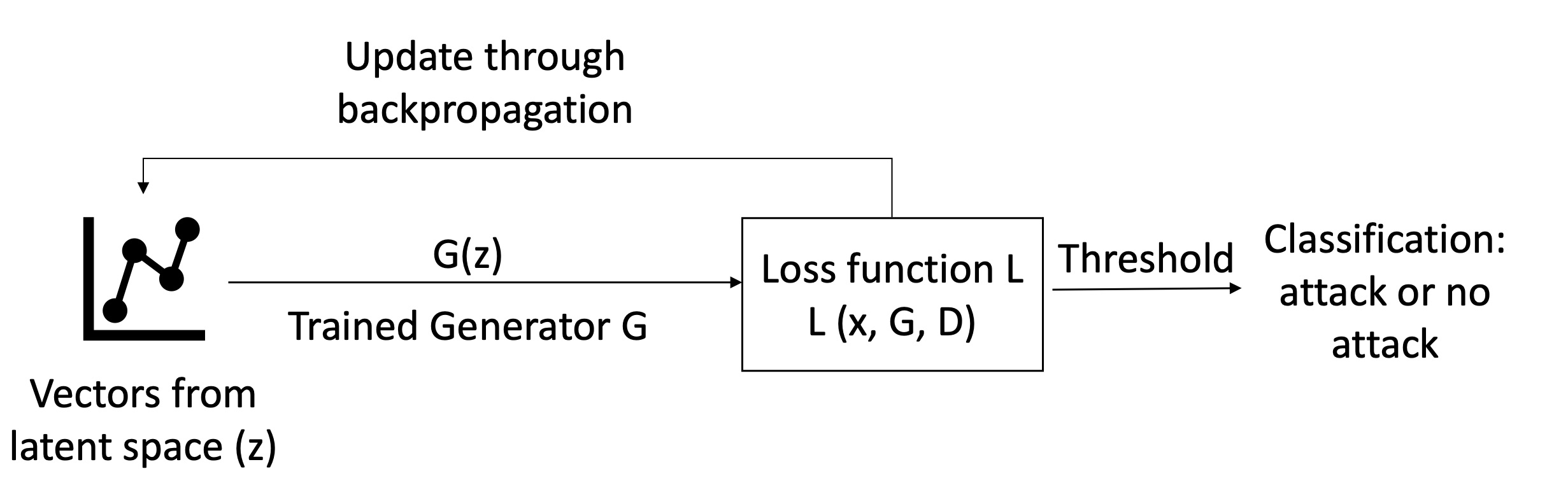}
    \caption{Demonstration of a GAN-based attack detection algorithm to measure detection score.}
    \label{fig:detection}
    \vskip-5pt
\end{figure}

It is worth noting that the GAN model does not have inverse mapping, i.e., $G\inv(\mb{x}) \in \mb{Z}: \mb{X} \rightarrow \mb{Z}$. Thus, to find the most similar $\mb{z}$ for a given data $\mb{x}$, which is a multivariate sample of vehicle speed, spacing, and acceleration, one must first obtain a random sample $\mb{z}_1 \in \mb{Z}$ and then feed it into the trained generator $G$ to generate a sample of {$G(\mb{z}_1)$}. Finally, an iterative back-propagation process is employed via $\gamma = 1,2,3,\cdots$ to find the most similar {$G(\mb{z}_\gamma)$}. This process is illustrated in Fig.~\ref{fig:detection}. 
There are two loss functions used during the process of mapping real data to the latent space from $\mb{x} \in \mb{X}$ to $\mb{z} \in \mb{Z}$, which are both updated through gradients for the parameters of $\mb{z}_1$ until the optimized $\mb{z}_\gamma$.

The first part of the loss function is the residual loss $L_R$, which measures the element-wise differences between two data sets, namely the real sample and the generated sample:
\begin{equation}\label{eq:loss_R}
 L_R = \sum\left|\mb{x} - G(\mb{z})\right|.
\end{equation}

\edit{The second part is the discrimination loss $L_D$, which measures the difference in feature representation between real and generated data.} It utilizes the learned feature representation $h(\mb{x}_i)$ of the convolutional layer from the discriminator, given by:
\begin{equation}\label{eq:loss_D}
L_D = \sum\left|h(\mb{x}) - h( G(\mb{z}))\right|.
\end{equation}

The final loss function $L$ is defined as the weighted sum of the residual loss $L_R$ and the discrimination loss $L_D$:
\begin{equation}\label{eq:loss}
 L = (1-\lambda) \cdot L_R + \lambda \cdot L_D.
\end{equation}

As demonstrated in Fig.~\ref{fig:outline}, the trained GAN model is used to determine whether a new observation is normal or attacked in the proposed detection framework. During the adversarial training process, the generator $G$ has learned both the unattacked data distribution $p_{\text{z}}(\mb{z})$ from the latent space $\mb{z} \in \mb{Z}$ and the unattacked data distribution $p_{\text{data}}(\mb{x})$ from the real data distribution $\mb{x} \in \mb{X}$. The generated data is based on normal traffic data distribution, since it is trained from unattacked trajectory data. The loss function $L$ shown in~\eqref{eq:loss} is employed to determine if new observations fit into the normal data distribution. A smaller loss indicates that the given data fits better with the normal traffic data learned by the generator $G$ in the adversarial training, while a larger loss reveals that the given data is from attacked traffic data. The corresponding threshold is determined from the observation of normal traffic trajectory data in the adversarial training. Any loss value greater than the threshold is considered as attacked traffic data. The parameter $\lambda$ is determined empirically following~\cite{bashar2020tanogan}.

The detection strategy is further illustrated in Fig.~\ref{fig:detection}. After training the GAN model with unattacked traffic data, one has the generator and discriminator models ready for use. In the first step of the detection algorithm, a threshold is determined such that loss values below it correspond to normal traffic conditions while those above it signify the occurrence of attacks. To find the threshold, we employ another dataset of unseen normal traffic data called validation data. The validation data differs from the training data used to train the GAN model. This data has a smaller sample size than the training data, specifically a ratio of 4:1 to the testing data is used to measure the model performance, a standard split ratio in data science experiments. As mentioned before, the GAN model does not have inverse mapping. Thus, finding the most similar $G(\mb{z}_\gamma)$ through an iterative process is time-consuming. In our approach, we {address} this issue using a smaller sample size of validation data to calculate the threshold. It shows {that an appropriate threshold can be obtained even from unseen normal trajectory data}.

\section{Numerical Results}\label{numerical}

In this section, we examine both the impacts of cyberattacks on individual vehicles and the broader influence of attacked ACC vehicles on traffic flow. Moreover, we evaluate the performance of the proposed detection model with a comparison to some machine learning based detection models recently developed.

\subsection{Impacts of Cyberattacks on Individual Vehicles}\label{section3}

We simulate cyberattacks introduced in the previous section to study their impacts on individual vehicles at a microscopic level. \editnew{Due to the lack of real-world ACC traffic flow data, synthetic simulations} are conducted with \texttt{MATLAB 2021a} on a computer with an Intel i7-9750 CPU @ 2.6 GHz processor. The three types of attacks are studied via a \editnew{200~m single-lane} ring-road experiment {which has been widely adopted in prior studies~\cite{giammarino2020traffic,gisolo2022nonlinear}.}
A ring road simulation is used since it represents infinite traffic flow, where each vehicle has a lead vehicle and a following vehicle, and thus allows for the natural development of traffic phenomena such as phantom traffic jams without being influenced by boundary conditions~\cite{Sugiyamaetal2008}.


As seen in the literature~\cite{talebpour2016influence,shang2021impacts,wang2023general}, the intelligent driver model (IDM)~\cite{treiber2000congested} with different model parameter values is widely used to differentiate the driving behavior of HVs and ACC vehicles, given by:
\begin{align}\label{eq:IDM}
    & f(\mathbf{\theta},s,v,\Delta v) = \alpha\left[1-\left(\frac{v}{v_d}\right)^\delta - \left(\frac{\hat{s}(v, \Delta v)}{s}\right)^2\right],
\end{align}
where,
\begin{align}
    \hat{s}(v, \Delta v) = \eta + \tau v - v \Delta v/(2\sqrt{\alpha \beta}).
\end{align}


The vector of model parameters, $\mathbf{\theta}$, is defined as $\mathbf{\theta} = [\alpha, \beta, \delta, \eta, \tau, v_d]^\top$, where $\alpha$ is the maximum acceleration, $\beta$ is the comfortable deceleration, $\delta$ is an acceleration exponent, $\eta$ is the jam distance, $\tau$ is the time gap, and $v_d$ is the desired speed. For ACC vehicles, the model parameter values are taken as $\mathbf{\theta_{\text{ACC}}} = [0.6, 5.2, 15.5, 6.3, 2.2, 44.1]^\top$ from~\cite{souza2019calibrating}, which shows that modeling ACC vehicles using the IDM can fit the ACC driving data well based on calibration using field experiments data of ACC vehicles~\cite{gunter2020are}. For HVs, the IDM parameter values are adopted from~\cite{kesting2008calibrating}, with $\mathbf{\theta_{\text{Human}}} = [1.06, 2, 4, 3.4, 1.26, 30]^\top$.

\begin{figure}[t!]
\centering
\subfloat[Speed]{\label{fig:b_norm}\includegraphics[width=0.25\textwidth]{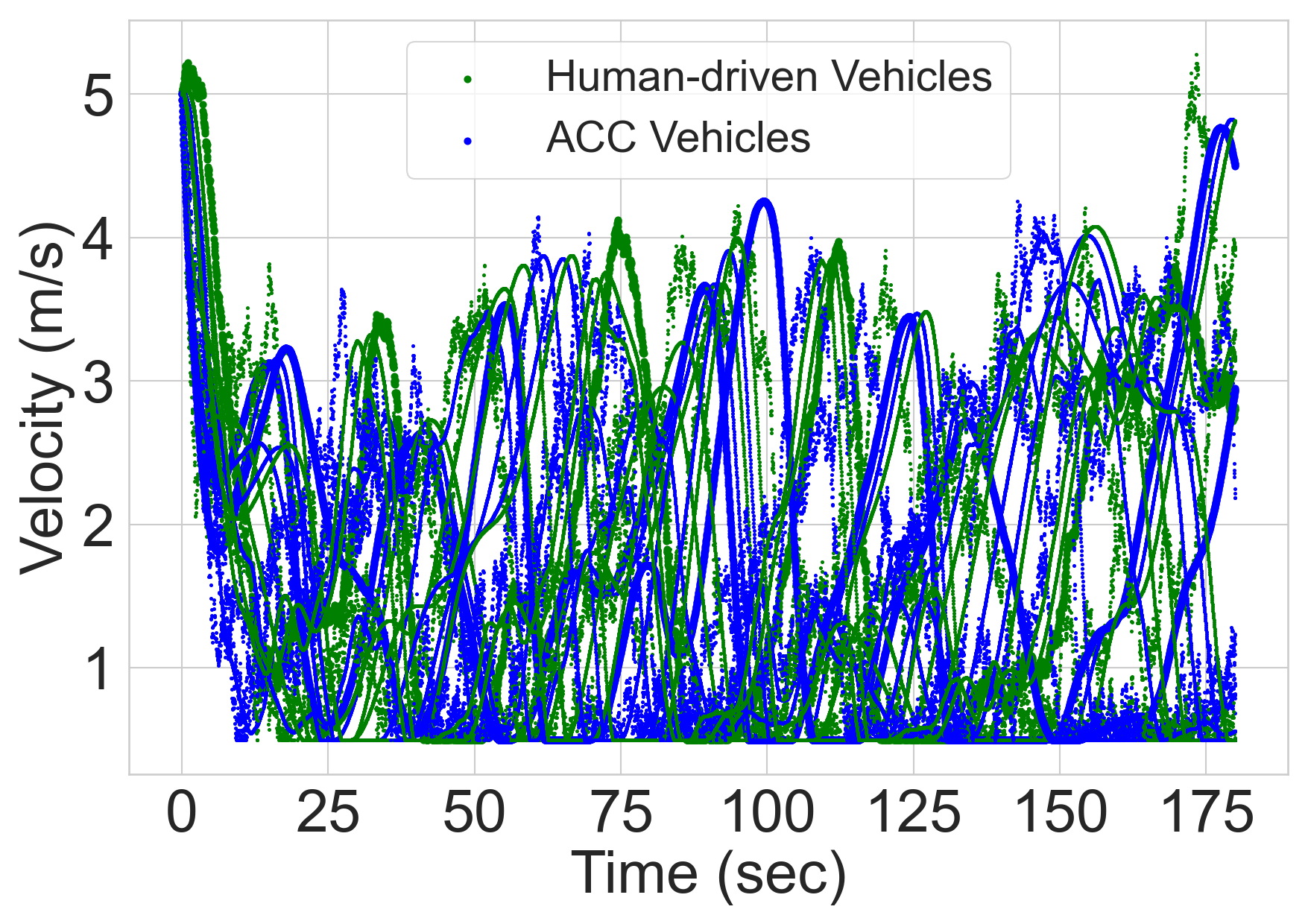}}
\subfloat[Position]{\label{fig:d_norm}\includegraphics[width=0.25\textwidth]{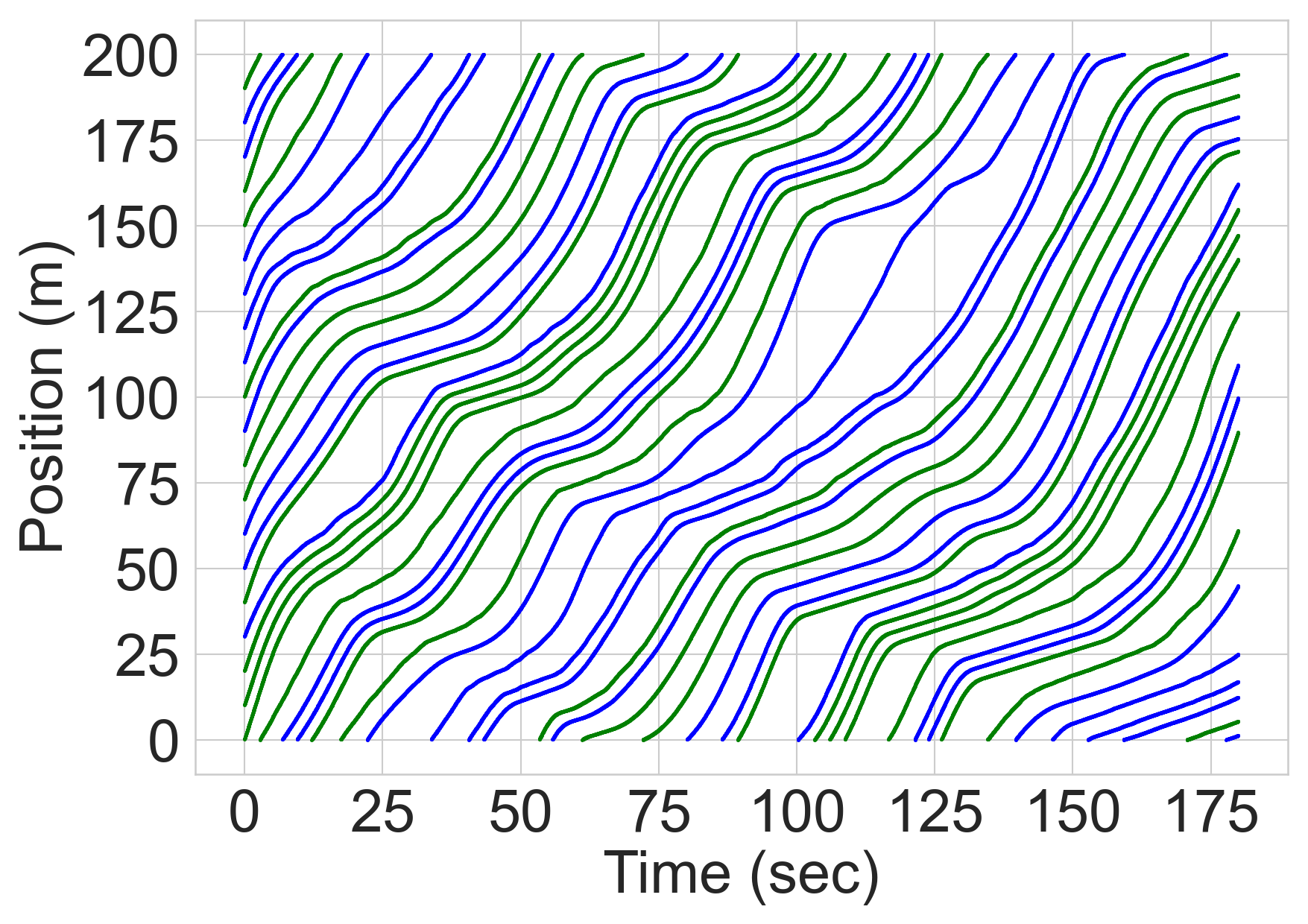}}%
\caption{\editnew{Trajectories of speed and position without attack for human-driven vehicles, in green; and ACC vehicles, in blue.}}
\label{fig:norm_sim}
\end{figure}

The initial inter-vehicle spacing between any two consecutive vehicles is set as the total length of the track divided by the number of vehicles.  In other words, vehicles are evenly distributed at the beginning of the simulation. {The initial velocity is set to zero, and a small perturbation of $\mathcal{N}(0, 0.5)$ is applied only to the first vehicle in the ring to initiate the simulation. } \edit{A sample profile of normal trajectory data without attack generated by the ring simulation is shown in Fig.~\ref{fig:norm_sim}, which contains profiles of the speed and position of all the simulated vehicles.} There are 20 vehicles in total, with 50\% being ACC vehicles \edit{as an example for illustration}. {Note that the traffic oscillations in Fig.~\ref{fig:norm_sim} result from the string unstable driving behaviors of both human drivers and ACC vehicles, as seen in~\cite{shang2021impacts}.} Among the 10 ACC vehicles, half {are} randomly chosen to be attacked by Type~\rom{1}, Type~\rom{2}, and Type~\rom{3} attacks for Scenario 1, Scenario 2, and Scenario 3, respectively, in this section. {Since the simulation experiments are run with various ACC market penetration rates,} we fix the attacked percentage for ACC vehicles. Thus, 50\% of the ACC vehicles are assumed to be attacked at whatever ACC market penetration rates under Type~\rom{1}, Type~\rom{2}, and Type~\rom{3} attacks. Each attack scenario is investigated independently, and only one type of attack is considered in each simulation. The trajectories of the following vehicles in the ring simulation are computed based on the kinematic dynamics given by:
\begin{equation}\label{eq: simulation process}
\begin{bmatrix}
s_k \\
v_k
\end{bmatrix}_{t+\Delta t} =
\begin{bmatrix}
s_k \\
v_k
\end{bmatrix}_t 
+ 
\begin{bmatrix}
v_{k-1}-v_k \\
a_k
\end{bmatrix}_t \Delta t,
\end{equation}
where $k~(1 < k \leq 20)$ is a vehicle index, the acceleration of vehicle $k$, i.e., $a_k$, is calculated using~\eqref{eq:IDM}, and a time step of $\Delta t = 0.033$~sec is used \editnew{with a sampling rate of 30 Hz.} Note that when $k=1$, the lead vehicle is Vehicle 20 in the ring simulation.

\subsubsection{Scenario 1: Malicious Attacks on Vehicle Control (Type~\rom{1} Attack)}

\begin{figure}[t!]
\centering
\subfloat[Speed]{\label{fig:b_type1}\includegraphics[width=0.25\textwidth]{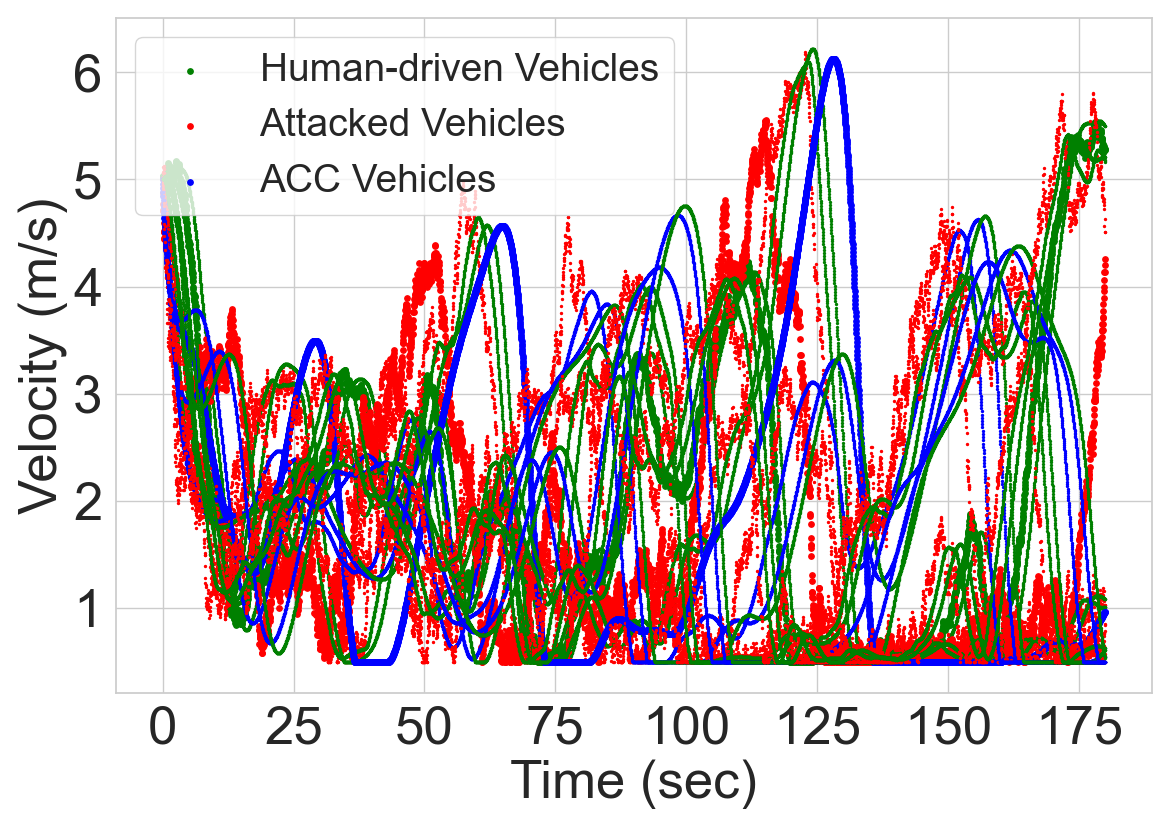}}
\subfloat[Position]{\label{fig:d_type1}\includegraphics[width=0.25\textwidth]{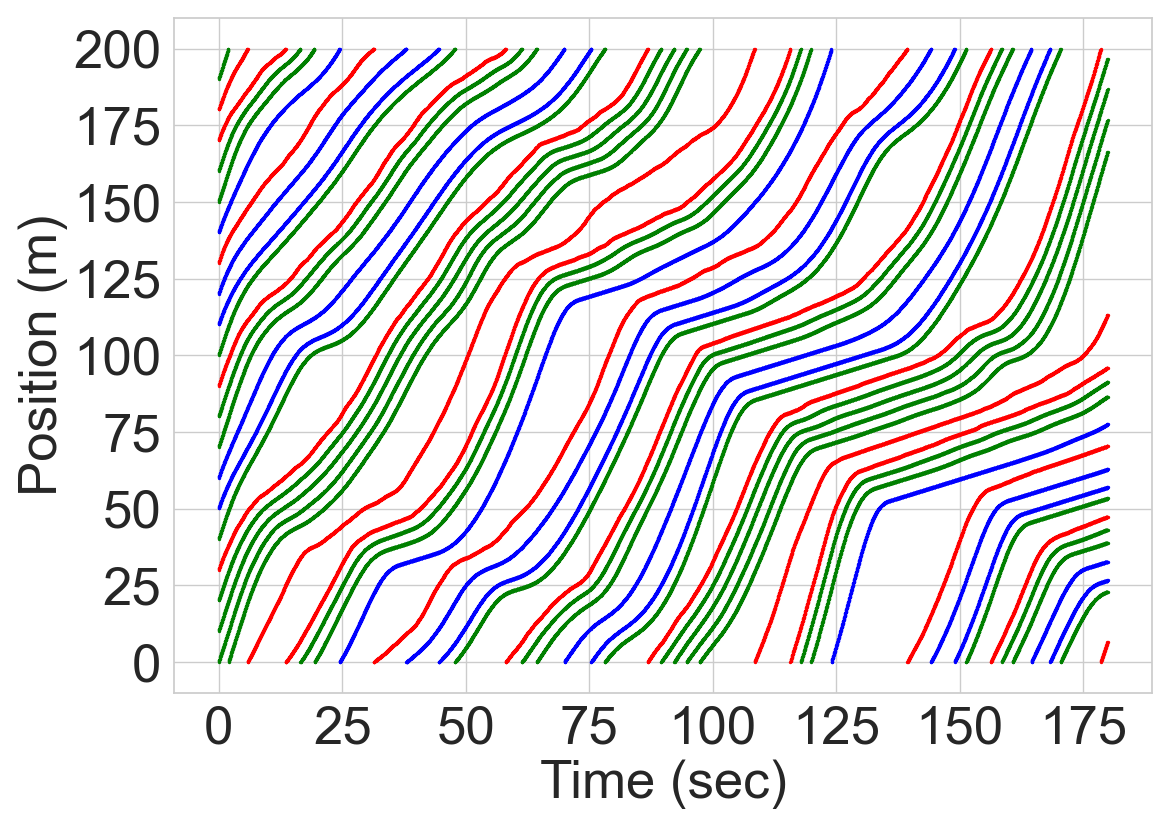}}%
\caption{\editnew{Trajectories of speed and position for ACC vehicles under Type~\rom{1} attack, in red; human-driven vehicles, in green; and unattacked ACC vehicles, in blue.}}
\label{fig:type_1_sim} 
\end{figure}

With a sample of normal traffic data shown in Fig.~\ref{fig:norm_sim}, we present the simulation results of the ring-road experiments considering Type~\rom{1} attacks on ACC vehicles in Fig.~\ref{fig:type_1_sim}. In this scenario, compromised vehicles are assumed to be attacked by a random attack, $\xi \sim \mathcal{N}(0,5)$ (a specific example for the purposes of illustration, with the unit of $\xi$ being the same as that of acceleration), directly acting on their acceleration. It is noted that the acceleration of any vehicle is physically bounded regardless of being attacked or not, as seen in~\eqref{eq:att1}. For demonstration, without loss of generality, attacks are assumed to follow a Gaussian distribution, while other forms of statistical distribution could also be considered~\cite{van2019real,wang2022planning}.

Compared with the normal traffic data presented in Fig.~\ref{fig:norm_sim}, the acceleration of the vehicle is being attacked, resulting in a smaller average speed for all the vehicles. The average spacing gap of attacked vehicles (Fig.~\ref{fig:d_type1}) is observed to be larger compared to its counterpart in the absence of attacks (Fig.~\ref{fig:d_norm}). Consequently, the position of vehicles deviates from their normal range, especially for attacked ACC vehicles, as shown in Fig.~\ref{fig:d_type1}.

\subsubsection{Scenario 2: False Data Injection Attacks on Sensors (Type~\rom{2} Attack)}

\begin{figure}[t!]
\centering
\subfloat[Speed]{\label{fig:b_type2}\includegraphics[width=0.25\textwidth]{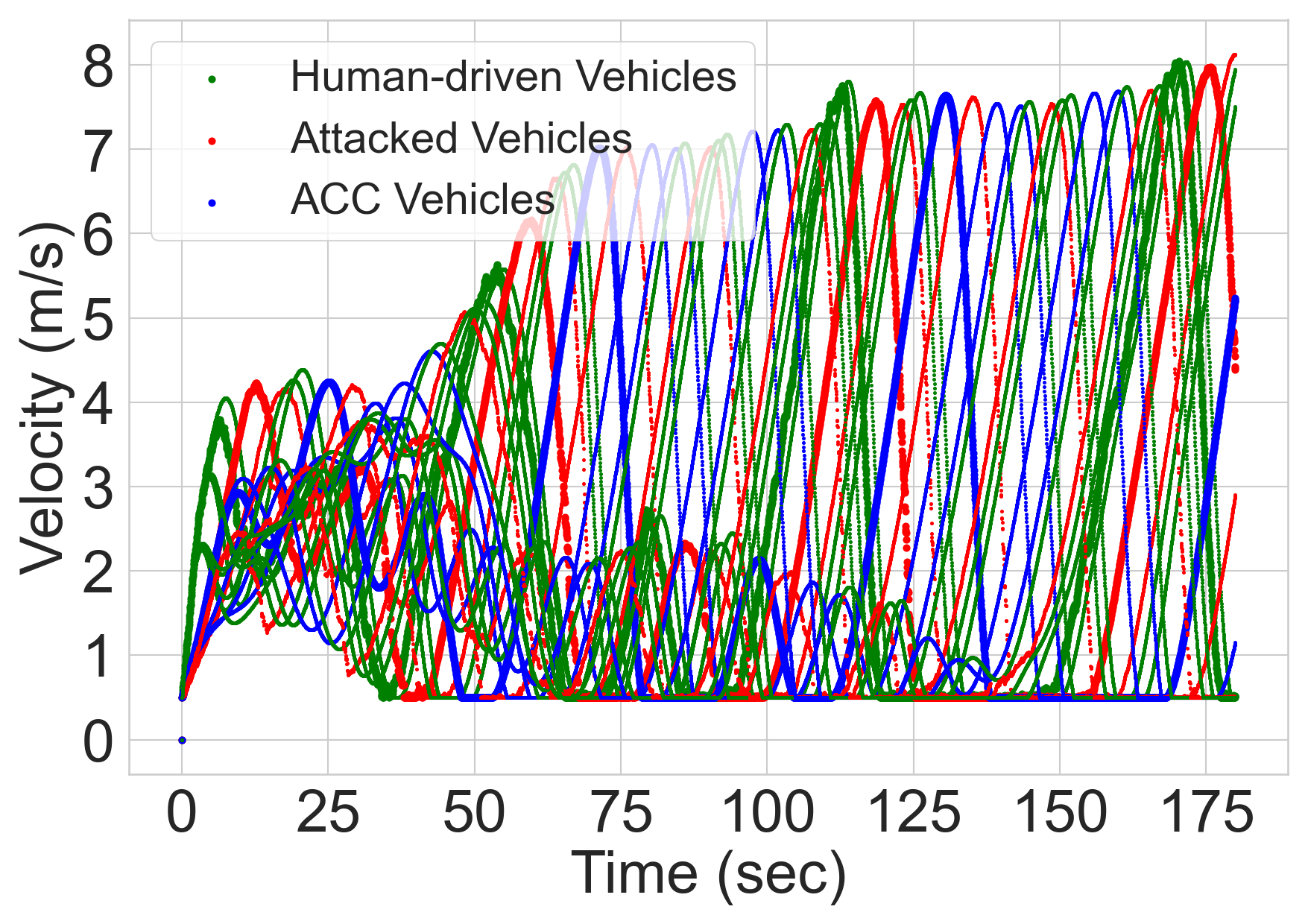}}
\subfloat[Position]{\label{fig:d_type2}\includegraphics[width=0.25\textwidth]{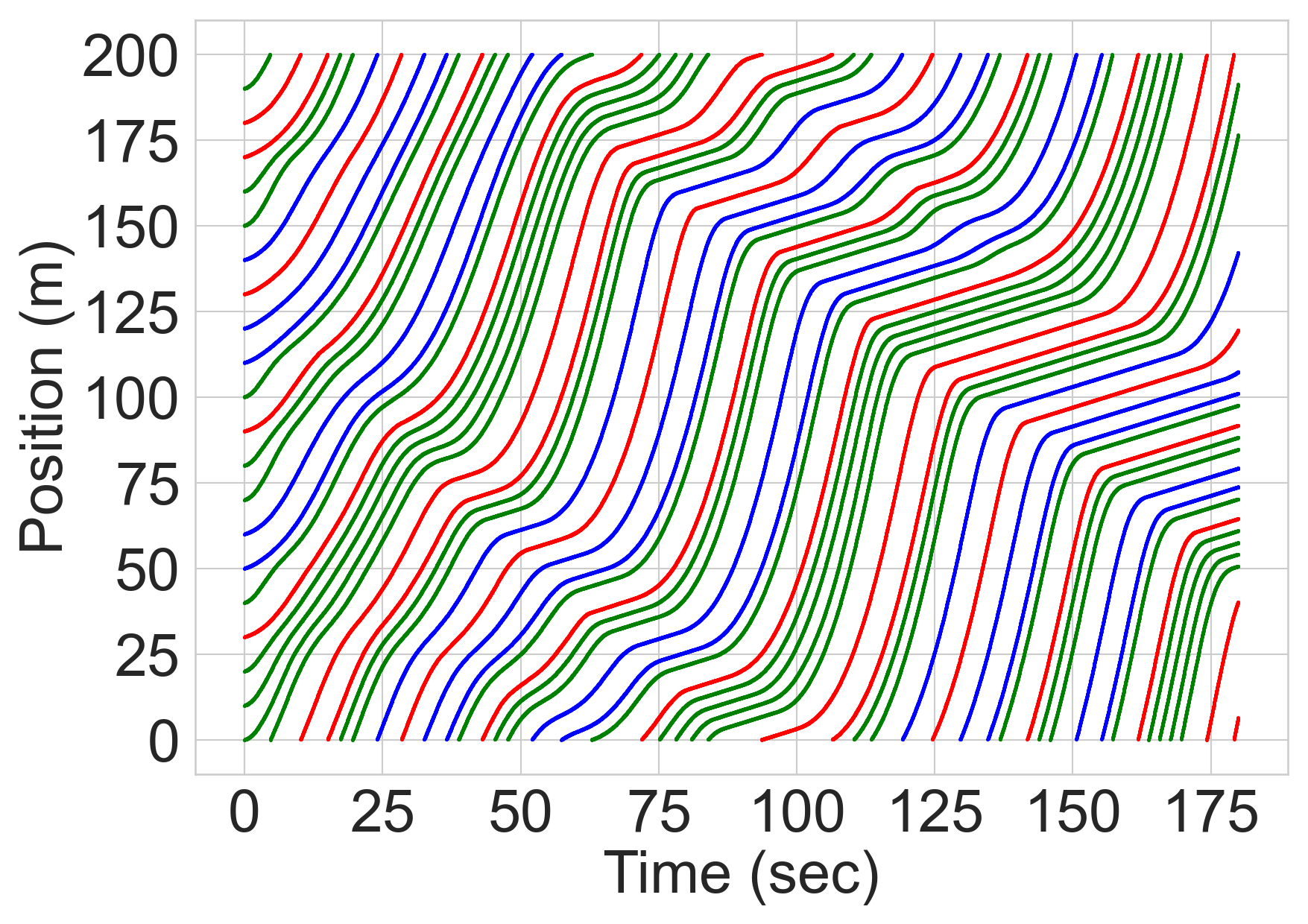}}%
\caption{\editnew{Trajectories of speed and position for ACC vehicles under Type~\rom{2} attack, in red; human-driven vehicles, in green; and unattacked ACC vehicles, in blue.}}
\label{fig:type_2_sim} 
\end{figure}

In this scenario, Type~\rom{2} attacks are assumed to be launched on vehicle sensor measurements. Specifically, false data, denoted by $\lambda_1$ and $\lambda_2$, are injected into the car-following measurements of $\mb{A} = [\Delta v, s]^\top$ as described by the first equation of~\eqref{eq:att2_2}. The false data injection attacks in this scenario, $\lambda_1$, and $\lambda_2$, are also assumed to follow a Gaussian distribution of $\mathcal{N}(0,5)$, similar to the previous scenario, with the units of $\lambda_1$ and $\lambda_2$ being same as those for spacing and relative speed, respectively. The rest of the simulation setting remains the same as in Scenario 1, with the same set of five ACC vehicles selected to be attacked by Type~\rom{2} attacks. The corresponding simulation results are shown in Fig.~\ref{fig:type_2_sim}. Compared with Type~\rom{1} attacks, this attack does not directly act on the control commands of vehicles. It is observed that vehicle speed, spacing gap, and position of Scenario 2 (Fig.~\ref{fig:type_2_sim}) are closer to the normal traffic data (Fig.~\ref{fig:norm_sim}), as opposed to those of Scenario 1 (Fig.~\ref{fig:type_1_sim}). This is because in Scenario 1 Type~\rom{1} attacks are directly launched on vehicle acceleration, resulting in a more severe impact on the driving behavior of attacked vehicles, compared to Scenario 2 where such impact is indirectly reflected through the first dynamic equation of~\eqref{eq:att2_2}.

\subsubsection{Scenario 3: Denial-of-Service (Type~\rom{3} Attack)}

\begin{figure}[t!]
\centering
\subfloat[Speed]{\label{fig:b_type3}\includegraphics[width=0.25\textwidth]{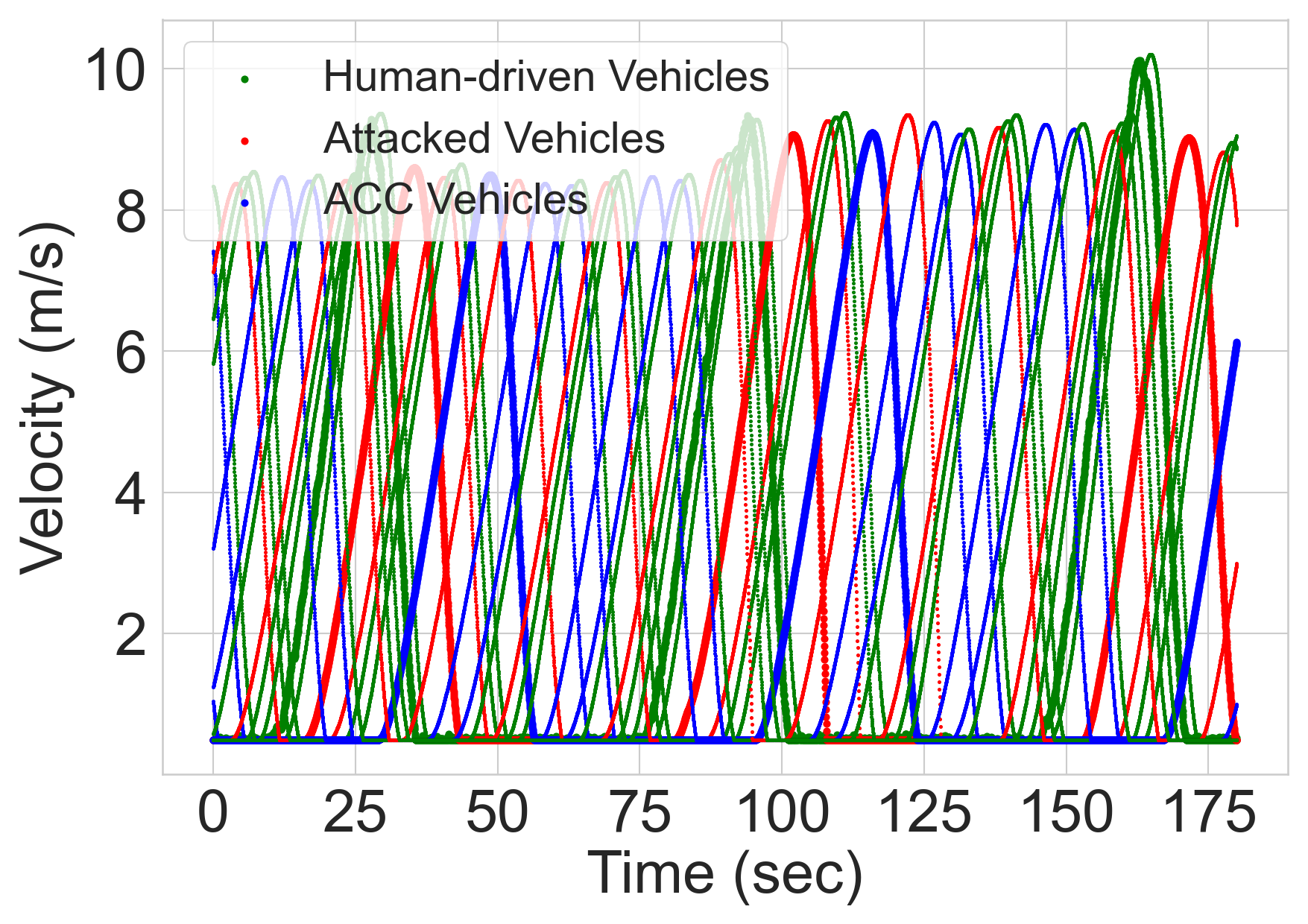}}
\subfloat[Position]{\label{fig:d_type3}\includegraphics[width=0.25\textwidth]{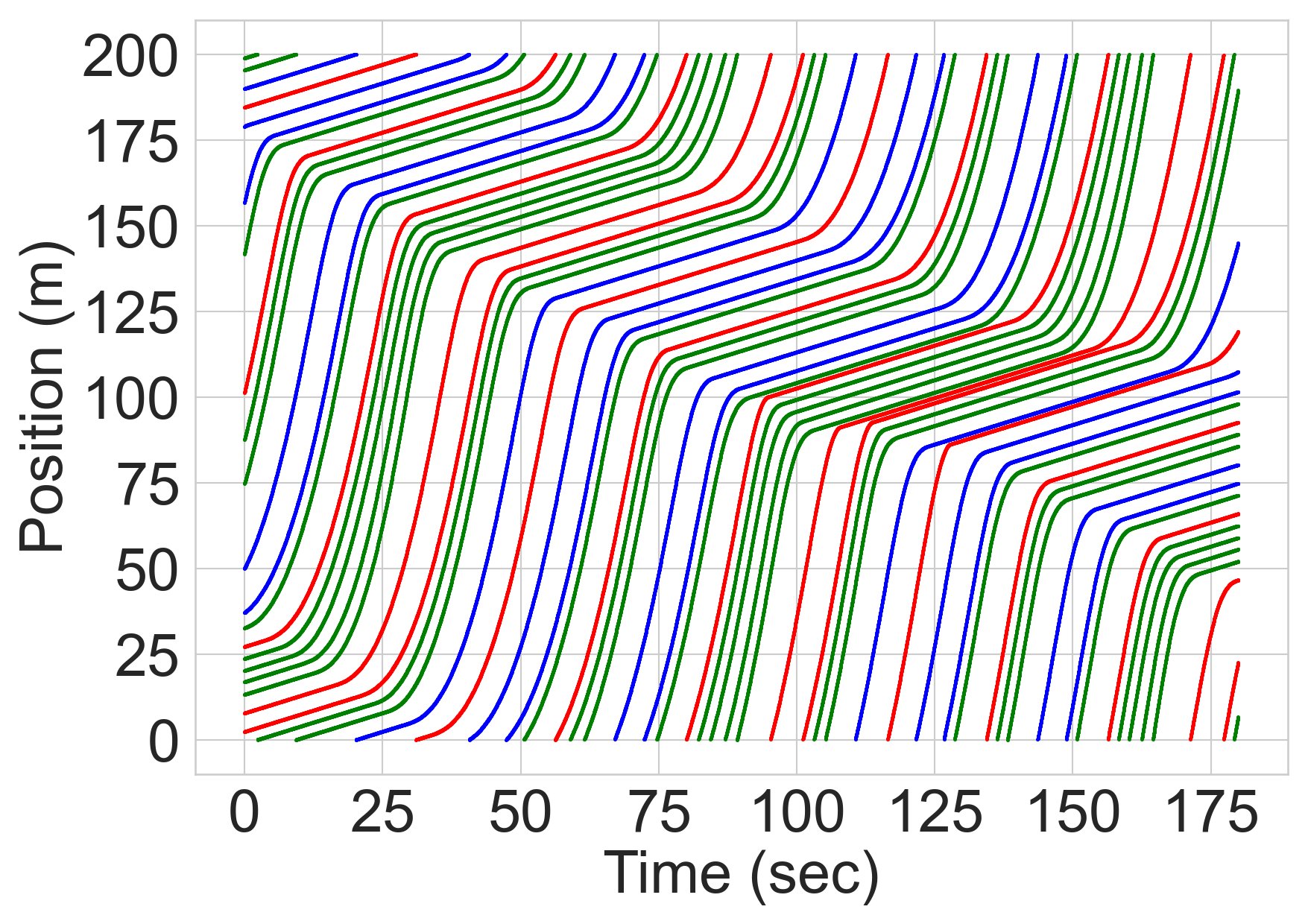}}%
\caption{\editnew{Trajectories of speed and position for ACC vehicles under Type~\rom{3} attack, in red; human-driven vehicles, in green; and unattacked ACC vehicles, in blue.}}


\label{fig:type_3_sim} 
\end{figure}

In this scenario, we consider DoS attacks on ACC vehicles, where a value of delay, $\omega$, is introduced to the communication of sensors in measuring car-following variables, i.e., $s$ and $\Delta v$, to mimic the effect of DoS attacks. Specifically, $\omega$ signifies the delay of data transmission in measuring $s$ and $\Delta v$ for the following vehicle. The corresponding results of this scenario are presented in Fig.~\ref{fig:type_3_sim}, with the same set of 5 ACC vehicles attacked. The middle part of the 50-second trajectory (80 s - 130 s) corresponds to the duration of DoS attacks. The delay in data transmission is assumed to be $\omega = 1$~sec for sensors to measure speed and spacing information from the preceding vehicle. While the speed of the attacked vehicles is similar to those of the normal traffic data, more traffic waves are observed in this scenario (Fig.~\ref{fig:d_type3}) compared to the normal scenario (Fig.~\ref{fig:d_norm}).

{In this study, we utilize various candidate attacks to demonstrate general trends regarding the influence of stealthy attacks on both individual driving behavior (microscopic) and overall traffic flow (macroscopic).} In view of the three scenarios considered above, it is observed that Type~\rom{1} attacks tend to have more severe impacts on the driving behavior of individual vehicles, compared to Type~\rom{2} and Type~\rom{3} attacks. This is consistent with the fact that Type~\rom{1} attacks directly act on vehicle acceleration while the others do not. {However, Type~\rom{2} and~\rom{3} attacks could result in vehicles driving very closely to each other with a small gap, thereby causing safety concerns.} In addition to the attacked ACC vehicles experiencing driving disturbances, other vehicles following the attacked vehicles also suffer from the resulting perturbations. In other words, attacks could have a broader impact on the bulk traffic, other than directly altering the driving behavior of individual vehicles.

\subsection{Impacts of Cyberattacks on Traffic Flow}\label{section4}

\begin{figure*}
\centering
\subfloat[][Scenario 1: no attack]{\includegraphics[width=.25\textwidth,height=0.16\textheight]{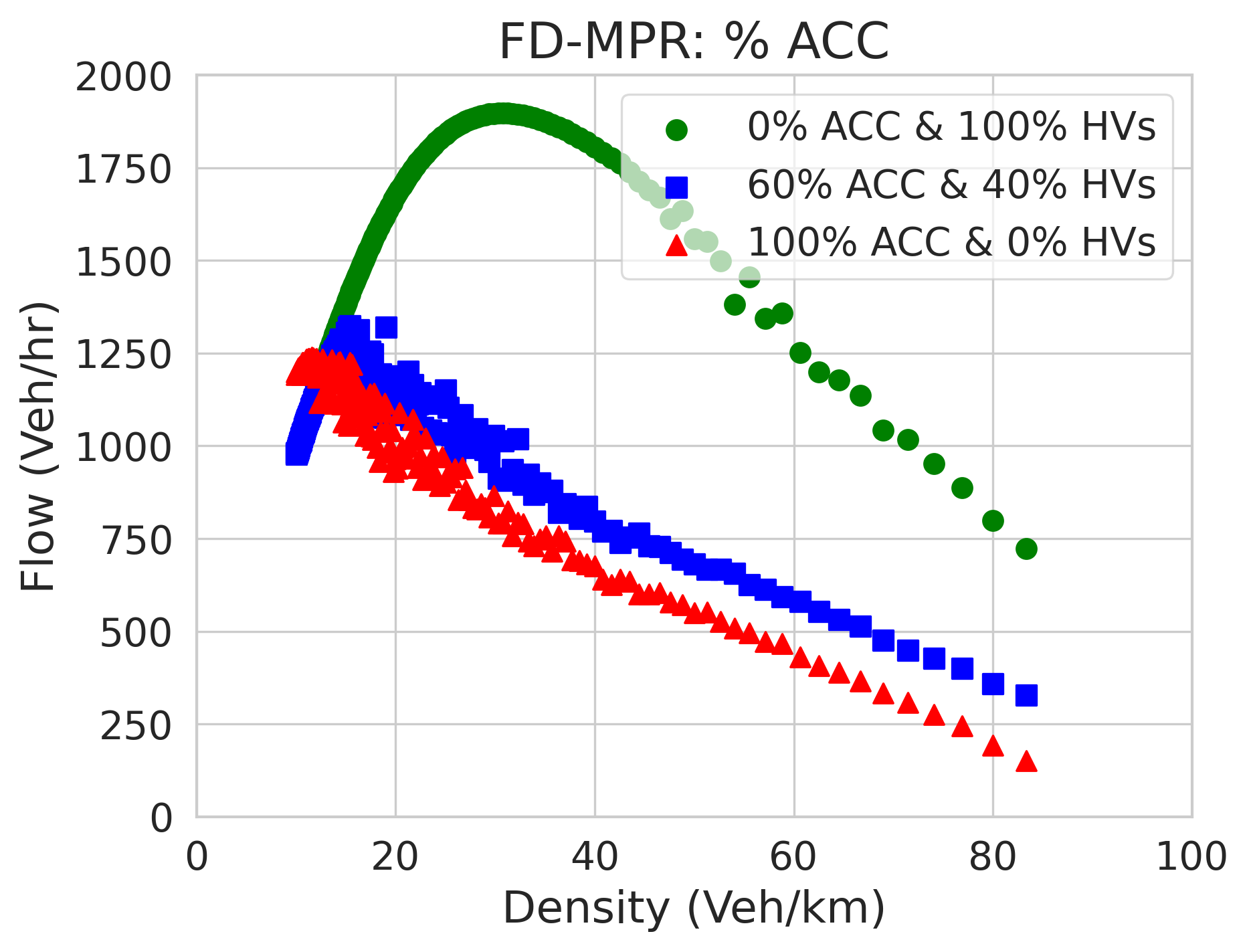}\label{fig:FD_norm}}
\subfloat[][Scenario 2: Type~\rom{1} attack]{\includegraphics[width=.25\textwidth,height=0.16\textheight]{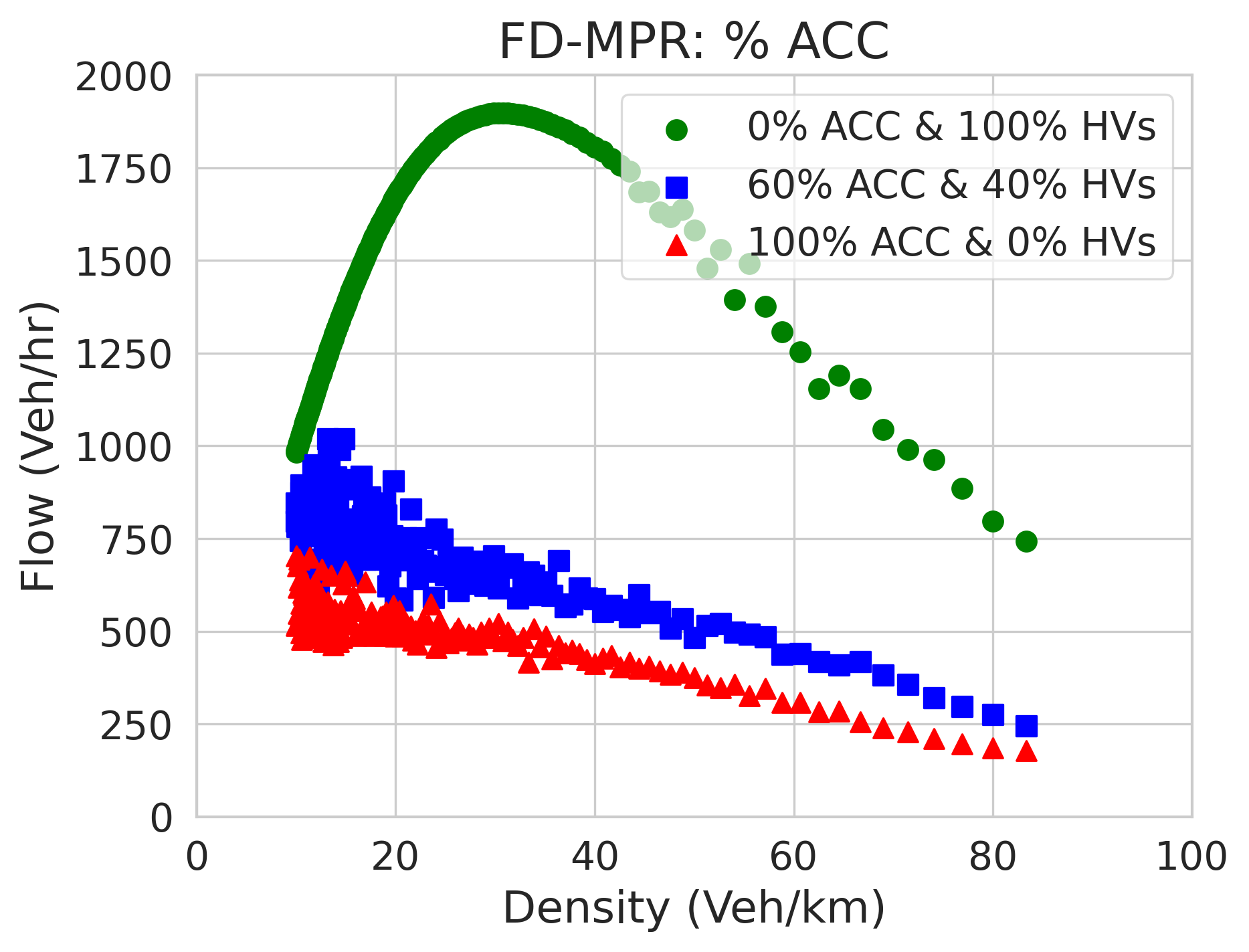}\label{fig:FD_type_1}}
\subfloat[][Scenario 3: Type~\rom{2} attack]{\includegraphics[width=.25\textwidth,height=0.16\textheight]{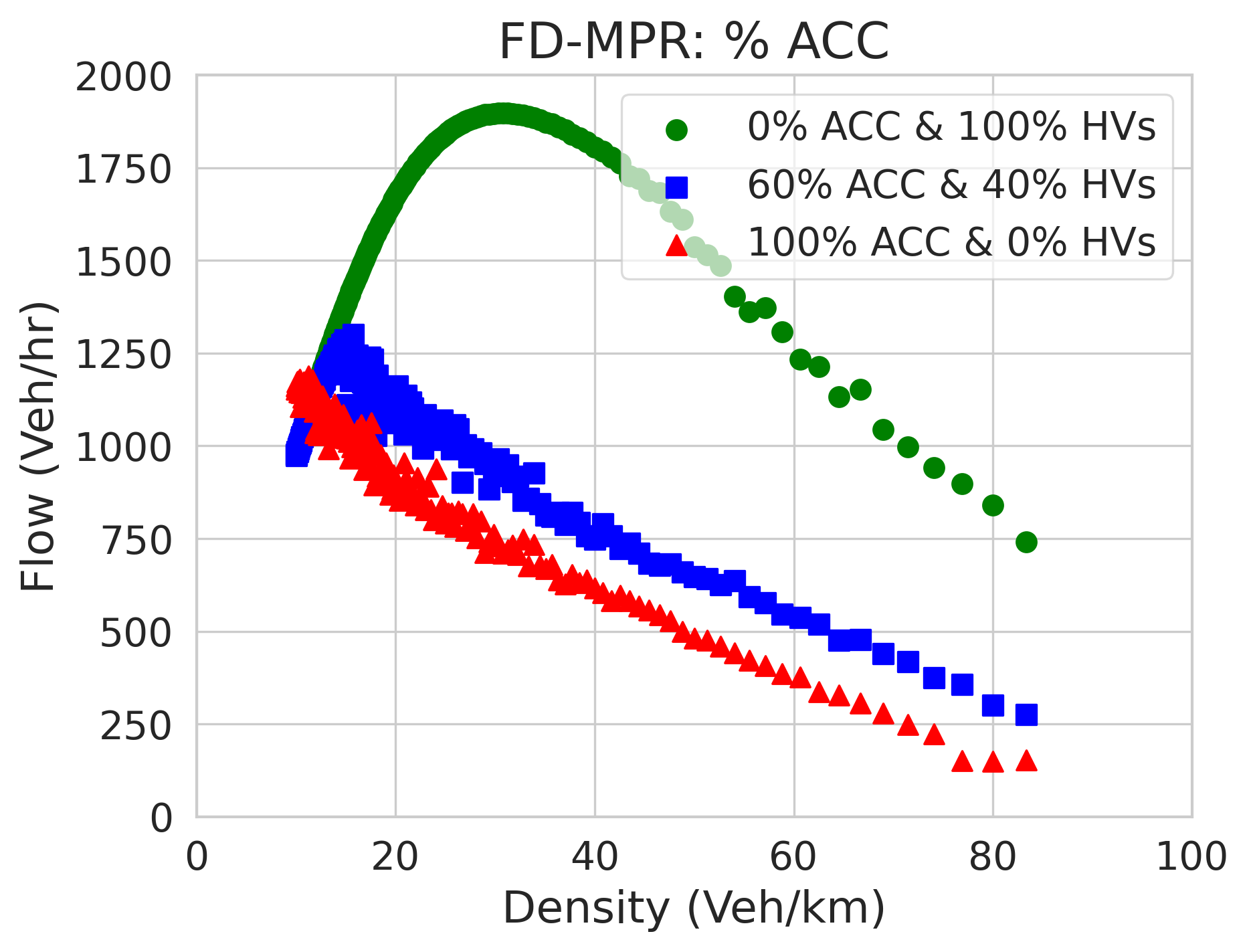}\label{fig:FD_type_2}}
\subfloat[][Scenario 4: Type~\rom{3} attack]{\includegraphics[width=.25\textwidth,height=0.16\textheight]{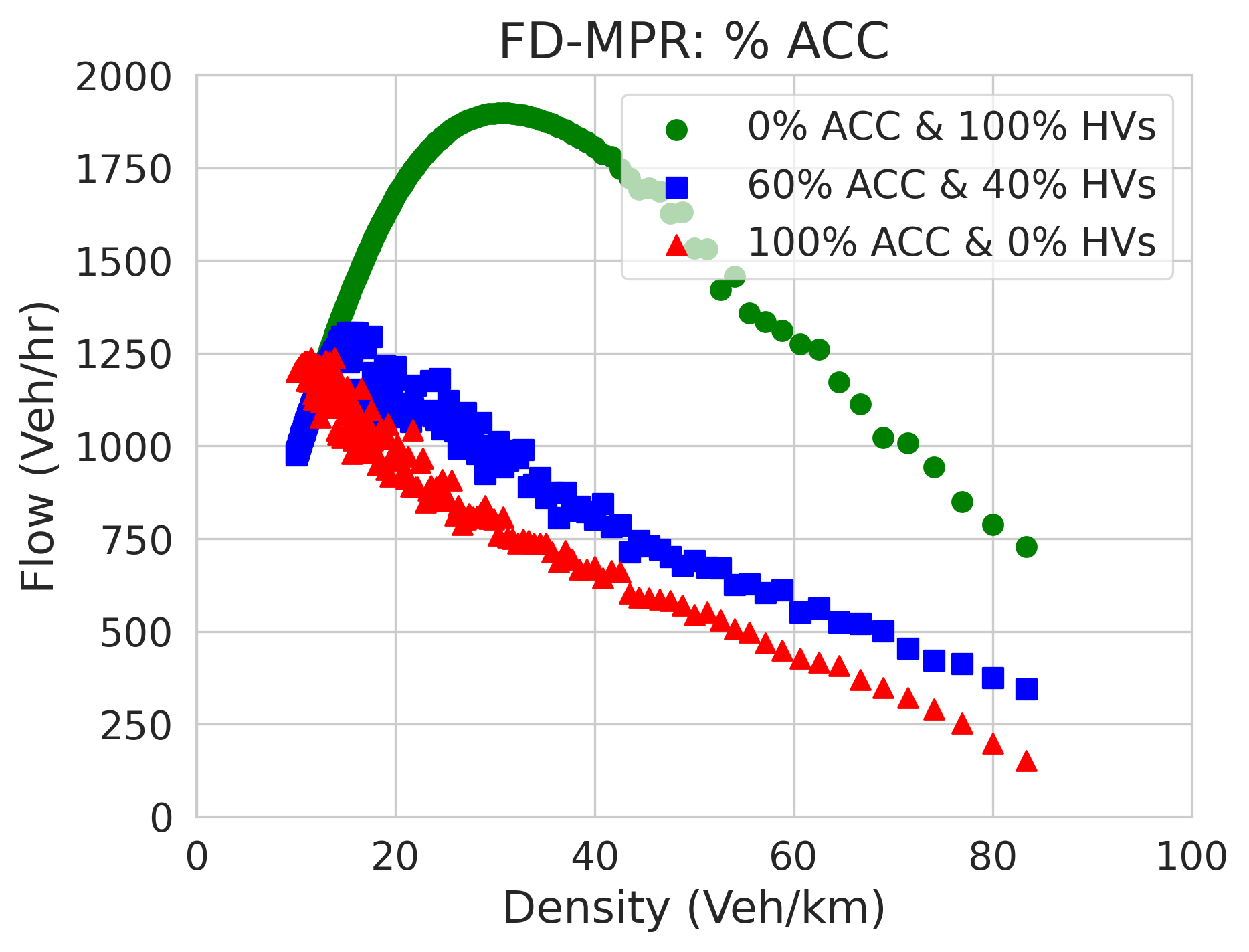}\label{fig:FD_type_3}}
\caption{Fundamental diagrams from simulating a ring of human-driven vehicles and commercially available ACC vehicles at different market penetration rates.}
\label{fig:free_flow}
\end{figure*}

Next, we consider the flow-level impact of the above attacks. To this end, we first need to recall the necessary traffic flow theory to inspect the results. Modeling traffic flow dynamics has been of interest in the transportation community for decades. Greenshields\etal~\cite{greenshields1934photographic} put the first effort into discovering the relationship between traffic flow and traffic density, \edit{known as a fundamental diagram (FD)}. {A series of studies have been conducted to investigate and obtain the fundamental diagram based on observation and collection of traffic flow data~\cite{greenshields1934photographic, coifman2015empirical}.}

As introduced above, a fundamental diagram \edit{characterizes} the relationship between traffic density and traffic flow (flux). To obtain the fundamental diagram we conduct multiple experiments to collect the traffic flow characteristics. {Following the setup of the ring simulation, we simulate 20 vehicles on a loop.} By changing the length of the loop, traffic density, \edit{denoted by} $\rho$, is also changed. While simulating the traffic flow, vehicle speeds are collected and the average speed ($\bar{v}$) is calculated. Consequently, the flow-density relationship is obtained as follows:
\begin{equation}
 q = \rho \bar{v},
\end{equation}
where $q$ is traffic flow. Hence, a fundamental diagram showing the flow-density relationship can be obtained through multiple tests by changing the loop length \edit{(equivalently changing traffic density)}.

In a fundamental diagram, we focus on the road capacity $Q$, the critical density $\rho_c$, and the jam density $\rho_j$. Typically, a fundamental diagram can be split into two regimes: the congested (car-following) regime and the uncongested (free-flow) regime. The congested regime ($\rho>\rho_c$) refers to the traffic flow with high density and low speed, \edit{whereas} the uncongested regime ($\rho<\rho_c$) \edit{corresponds to that} with low density and high speed. The highest traffic flow rate is the road capacity ($Q$) \edit{with the corresponding density called the critical density $\rho_c$.}


\subsubsection{Traffic Fundamental Diagram without Attack}
Fig.~\ref{fig:FD_norm} shows fundamental diagrams for the mixed autonomy traffic flow with commercially available ACC vehicles at different market penetration rates (MPRs) ranging from 0\% to 100\% without attack. It is observed that the capacity decreases from around 1,900 (veh/hr) to {1,250}(veh/hr) as the MPR increases \edit{from 0\% to 100\%}. This is consistent with the findings of~\cite{shang2021impacts} \edit{showing that capacity decreases with the increase of the MPR of commercially available ACC vehicles, since the headway of such vehicles is increased for safety purposes as opposed to HVs. When the MPR increases, the critical density tends to decrease with more scatter observed in the congested regime of the fundamental diagram, as seen in Fig.~\ref{fig:FD_norm}, indicating more oscillations in the traffic flow.}

\subsubsection{Impacts of Cyberattacks on Traffic Fundamental Diagram}

\edit{In this section, we present numerical results on fundamental diagrams with ACC vehicles being attacked under the three types of attacks introduced before. For comparison with the case without attacks (Fig.~\ref{fig:FD_norm}), we show fundamental diagrams at the ACC MPR of 0\%, 60\%, and 100\%.}


Fig.~\ref{fig:FD_type_1} shows the fundamental diagrams corresponding to Scenario 1. Since attacks are launched only on ACC vehicles, \edit{the fundamental diagram shown in Fig.~\ref{fig:FD_type_1} remains the same as Fig.~\ref{fig:FD_norm}, at 0\% MPR of ACC vehicles. However, at 60\% MPR the traffic capacity is significantly reduced from around {1,900} (veh/hr) (Fig.~\ref{fig:FD_norm}) to {1,100}(veh/hr) (Fig.~\ref{fig:FD_type_1}). A similar reduction in capacity is also observed at 100\% MPR, from {1,100 (veh/hr)} to {750} (veh/hr), as shown in comparing Fig.~\ref{fig:FD_norm} with Fig.~\ref{fig:FD_type_1}. This is because malicious attacks on vehicle acceleration alter individual vehicle driving behavior and cause disturbances to traffic flow, thereby resulting in an increased average headway and hence a decrease in traffic capacity.} 


Fig.~\ref{fig:FD_type_2} shows the fundamental diagrams for Scenario 2. As in Scenario 1, the fundamental diagram at 0\% MPR is the same as Fig.~\ref{fig:FD_norm} (without attack). The fundamental diagram for the MPR of 60\% in this case is similar to the normal scenario {shown} in Fig.~\ref{fig:FD_norm}. The capacity $Q$ and the shape of the fundamental diagram are also similar to the scenario in the absence of attacks. This is consistent with the findings of Section~\ref{section3} concerning impacts of Type~\rom{2} attacks on individual vehicles, where the speed, spacing gap, and position are similar to the scenario of normal traffic. Consequently, the resulting fundamental diagram does not differ much from normal.

\edit{Fig.~\ref{fig:FD_type_3} shows the fundamental diagrams for Scenario 3. In the absence of attacks, the fundamental diagram at 0\% MPR (Fig.~\ref{fig:FD_type_3}) is the same as the normal case without attack (Fig.~\ref{fig:FD_norm}). The fundamental diagram for the MPR of 60\% is shown in Fig.~\ref{fig:FD_type_3}, which is also similar to that of Fig.~\ref{fig:FD_norm} for the same reason as in Scenario 2.}


\edit{Given the three scenarios above, it is observed that the fundamental diagram under Type~\rom{1} attacks tends to be impacted more severely than the cases of Type~\rom{2} and Type~\rom{3} attacks. This is because Type~\rom{1} attacks act directly on vehicle acceleration, causing more significant impacts on the driving behavior of individual vehicles, compared to Type~\rom{2} and Type~\rom{3} attacks. However, {Type~\rom{2} and~\rom{3}} attacks could result in vehicles driving very closely to each other with a small gap, causing safety concerns, as discussed in {Fig.~\ref{fig:d_type2} and Fig.~\ref{fig:d_type3}}.}

\subsection{Performance Evaluation of the Detection Model}



\begin{figure}[t!]
    \centering
    \includegraphics[scale = 0.2]{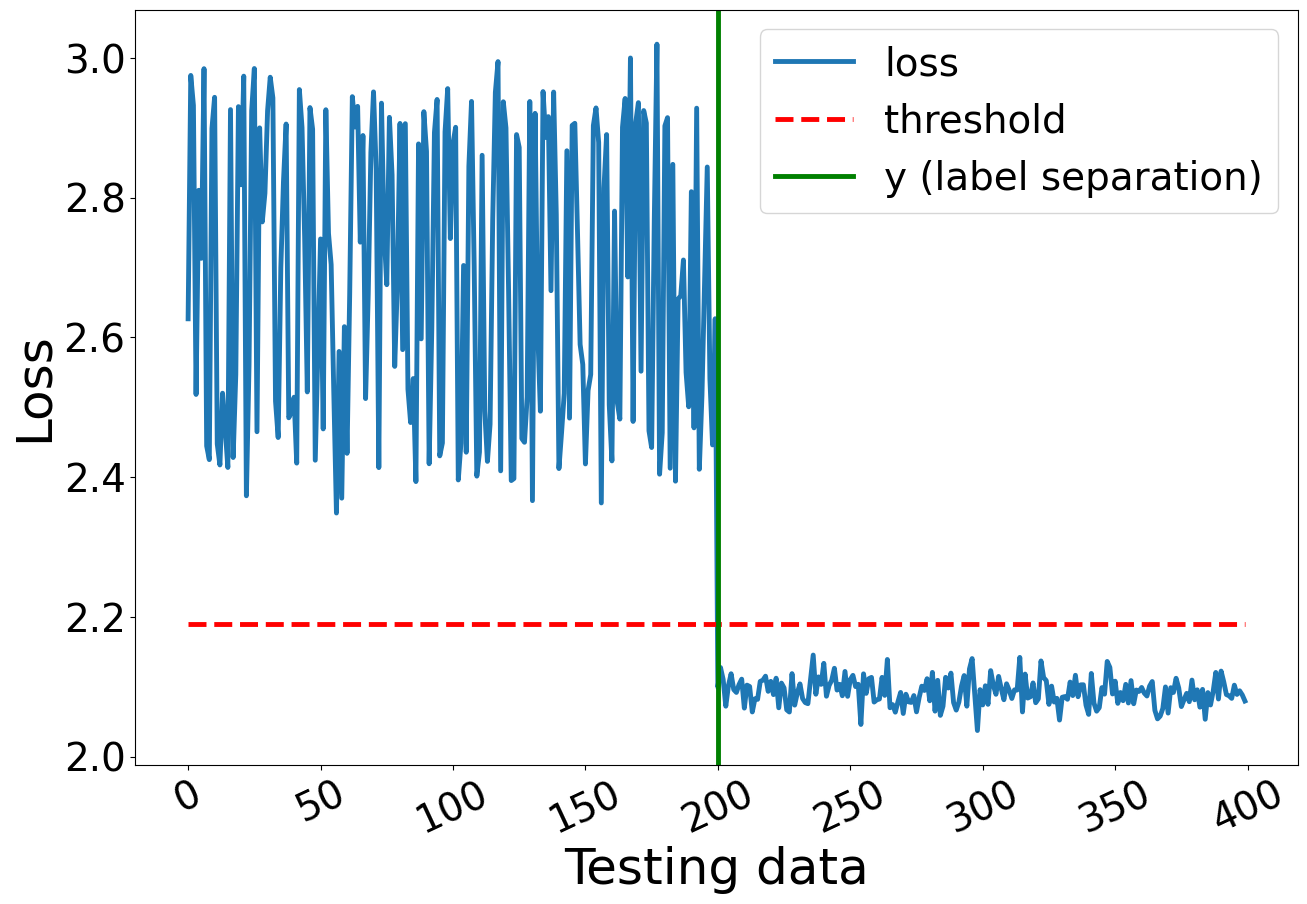}
    \caption{A sample model detection result of normal and attacked traffic data, where the left side is attacked data {resulting from Type~\rom{1} attacks}, and the right side is normal (unattacked) data. }
    \label{fig:detection_sample}
    \vskip-15pt
\end{figure}

The deep learning-based models are trained and tested on a single NVIDIA GeForce RTX 2070 and implemented in \texttt{PyTorch 1.11}. The model is trained using normal traffic data of vehicle speed, spacing, and acceleration, i.e., $[v, s, a]^\top$, and tested with unseen normal and attacked traffic data. A sample testing result is presented in Fig.~\ref{fig:detection_sample}, where the label $y = 0$ indicates normal traffic data and $y = 1$ signifies attacked traffic data {(Type~\rom{1} attack)}. The first 200 test data points are unseen attacked traffic data, and the last 200 data points are unseen normal traffic data. While some attacked data have a loss value lower than the threshold, it is observed that the proposed model distinguishes normal traffic data from hacked data fairly well. {It is worth noting that thousands of data points are used for the model performance evaluation. Fig.~\ref{fig:detection_sample} presents an example illustrating the loss values, wherein normal and attacked trajectories are clearly distinguished.}

{We also examine model performance for the three scenarios introduced in Section~\ref{section4} with different lengths of input data. All three types of attacks are mathematically defined in Section~\ref{section2}.} {We use these specific examples for illustration purposes only, while the proposed approach also works for other attacks.}


\subsubsection{Model Evaluation}

We use standard statistical metrics, namely \textit{Accuracy}, \textit{Precision}, \textit{Recall}, and $F_1$ score to evaluate the performance of the proposed detection model. Precision is the number of correctly detected positive instances divided by the number of total positive classifications the model makes. Recall is the number of correctly detected positive instances divided by the total positive instances in the model. Accuracy is calculated as the sum of true positives and true negatives divided by the total population. $F_1$ score is the harmonic mean of the precision and recall, with an equal weight given to precision and recall. A more detailed description of these statistical measures can be found in~\cite{hastie2009elements}. In the context of this study, the precision \edit{represents} the rate of successfully detecting real attacks. The recall measures the model's capability to detect positive (attacked) samples. \edit{It indicates the number of positive (attacked) samples correctly detected by the model and is only dependent on positive (attacked) samples.} 





\begin{table}[t!]
\vspace{10pt}
    \caption{Evaluation of Model Performance}
    \begin{center}
    \setlength\tabcolsep{3.5pt}
    \begin{tabular}{c c c c c c}
        \toprule
       \bf \multirow{2}{*}{\makecell{Attack \\ Type}} & \bf \multirow{2}{*}{\makecell{Input \\ Length (s)}} & \bf \multirow{2}{*}{\makecell{Accuracy}} & \bf \multirow{2}{*}{\makecell{Precision}} & \bf \multirow{2}{*}{\makecell{Recall}} & \bf \multirow{2}{*}{\makecell{$F_1$ Score}}  \\ \\
        \hline
         \multirow{6}{*}{\makecell{Type~\rom{1} \\ (Scenario 1)}} 
        &  \multirow{1}{*}{2} &  \multirow{1}{*}{86\%}  & \multirow{1}{*}{0.78}  & \multirow{1}{*}{1.00}   & \multirow{1}{*}{0.88} \\
        
      &  \multirow{1}{*}{4} &  \multirow{1}{*}{88\%}  & \multirow{1}{*}{0.80}  & \multirow{1}{*}{1.00}   & \multirow{1}{*}{0.89} \\
      
      &  \multirow{1}{*}{6} &  \multirow{1}{*}{91\%}  & \multirow{1}{*}{0.85}  & \multirow{1}{*}{1.00}   & \multirow{1}{*}{0.92} \\
     
      &  \multirow{1}{*}{8} &  \multirow{1}{*}{86\%}  & \multirow{1}{*}{0.77}  & \multirow{1}{*}{1.00}   & \multirow{1}{*}{0.87} \\
                      
      &  \multirow{1}{*}{10} &  \multirow{1}{*}{85\%}  & \multirow{1}{*}{0.77}  & \multirow{1}{*}{1.00}   & \multirow{1}{*}{0.87} \\
     
       &  \multirow{1}{*}{12} &  \multirow{1}{*}{88\%}  & \multirow{1}{*}{0.81}  & \multirow{1}{*}{1.00}   & \multirow{1}{*}{0.89} \\
          \hline
        \multirow{6}{*}{\makecell{Type~\rom{2} \\ (Scenario 2)}} 
        &  \multirow{1}{*}{2} &  \multirow{1}{*}{86\%}  & \multirow{1}{*}{0.78}  & \multirow{1}{*}{1.00}   & \multirow{1}{*}{0.88} \\
        
      &  \multirow{1}{*}{4} &  \multirow{1}{*}{88\%}  & \multirow{1}{*}{0.8}  & \multirow{1}{*}{1.00}   & \multirow{1}{*}{0.89} \\
      
      &  \multirow{1}{*}{6} &  \multirow{1}{*}{91\%}  & \multirow{1}{*}{0.85}  & \multirow{1}{*}{1.00}   & \multirow{1}{*}{0.92} \\

         &  \multirow{1}{*}{8} &  \multirow{1}{*}{84\%}  & \multirow{1}{*}{0.75}  & \multirow{1}{*}{1.00}   & \multirow{1}{*}{0.86} \\
         
      &  \multirow{1}{*}{10} &  \multirow{1}{*}{85\%}  & \multirow{1}{*}{0.77}  & \multirow{1}{*}{1.00}   & \multirow{1}{*}{0.87} \\
     
  &  \multirow{1}{*}{12} &  \multirow{1}{*}{88\%}  & \multirow{1}{*}{0.81}  & \multirow{1}{*}{1.00}   & \multirow{1}{*}{0.89}\\
          \hline
          
         \multirow{6}{*}{\makecell{Type~\rom{3} \\ (Scenario 3)}} 
        &  \multirow{1}{*}{2} &  \multirow{1}{*}{86\%}  & \multirow{1}{*}{0.78}  & \multirow{1}{*}{1.00}   & \multirow{1}{*}{0.88} \\
        
      &  \multirow{1}{*}{4} &  \multirow{1}{*}{88\%}  & \multirow{1}{*}{0.80}  & \multirow{1}{*}{1.00}   & \multirow{1}{*}{0.89} \\
      
      &  \multirow{1}{*}{6} &  \multirow{1}{*}{91\%}  & \multirow{1}{*}{0.85}  & \multirow{1}{*}{1.00}   & \multirow{1}{*}{0.92} \\
     
      &  \multirow{1}{*}{8} &  \multirow{1}{*}{86\%}  & \multirow{1}{*}{0.77}  & \multirow{1}{*}{1.00}   & \multirow{1}{*}{0.87} \\
                      
      &  \multirow{1}{*}{10} &  \multirow{1}{*}{85\%}  & \multirow{1}{*}{0.77}  & \multirow{1}{*}{1.00}   & \multirow{1}{*}{0.87} \\
     
       &  \multirow{1}{*}{12} &  \multirow{1}{*}{88\%}  & \multirow{1}{*}{0.81}  & \multirow{1}{*}{1.00}   & \multirow{1}{*}{0.89} \\
      
          \bottomrule
          
    \end{tabular}
    \end{center}
    \label{tab:stat}
    \vskip-10pt
\end{table}

Table~\ref{tab:stat} summarizes the model performance of the detection experiments with different lengths of input data. Observe that the model performs similarly well across the three scenarios. It also shows that model accuracy is not sensitive to the increase in the input data length. The experiments show that the proposed model can \edit{effectively} detect abnormal traffic with only 2 seconds of observed data. Generally, higher values of accuracy and $F_1$ score indicate better model performance in detecting attacked traffic. Note that the model has a lower precision than recall (recall is 1 in most of the experiments), meaning that it can \edit{detect all the three types of attacks} at various input lengths. However, lower precision indicates that the model could misclassify some normal traffic as \edit{being attacked}.


\subsubsection{Model Comparison}

Here we compare the proposed model with other neural network models that have been commonly applied to detect anomaly driving behavior of ACC vehicles from trajectory data, such as Autoencoder~\cite{gunter2021compromised} and Long Short-Term Memory (LSTM) neural network based GAN model~\cite{li2022detecting}. The comparison results are summarized in Table~\ref{tab:models}. It is observed that the proposed model \edit{outperforms the other models under all the scenarios considered.} Moreover, the other two models have a much lower precision value than the proposed model, indicating that they could misclassify more normal trajectories as \edit{attacked ones}.

Unlike many anomaly detection algorithms that have been applied \edit{to} time-series data, which mainly detect point anomalies (extremum), the proposed model is designed and customized for multi-dimensional vehicle trajectory data. Our model evaluates attacked (anomaly) trajectory using car-following sequence data based on domain knowledge and physical phenomena. \edit{Due to the unique structure of the trajectory data used in this study, one can not apply some popular machine learning based algorithms, such as k-Nearest Neighbors (kNN), identifying density-based local outliers (LOF), and Isolation Forest (IF), since they are intended for tabular data~\cite{zhao2019pyod,schmidl2022anomaly}. Moreover, many of those algorithms can only support univariate data consisting of only one variable. Hence, we can not compare our model with some time-series anomaly detection models proposed in the anomaly detection domain~\cite{zhao2019pyod,schmidl2022anomaly}.}

\begin{table}[t!]
\vspace{3pt}
    \caption{Model Comparison}
    \begin{center}
\setlength\tabcolsep{2.5pt}
    \begin{tabular}{c c c c c c}
        \toprule
        \bf Model & \bf Scenario & \bf Accuracy & \bf Precision &\bf Recall & \bf $F_1$ Score  \\
        \hline
         \multirow{3}{*}{\makecell{This paper}} 
        &  \multirow{1}{*}{ 1} &  \multirow{1}{*}{86\%}  & \multirow{1}{*}{0.78}  & \multirow{1}{*}{1.00}   & \multirow{1}{*}{0.88} \\
        
      &  \multirow{1}{*}{ 2} &  \multirow{1}{*}{86\%}  & \multirow{1}{*}{0.78}  & \multirow{1}{*}{1.00}   & \multirow{1}{*}{0.88} \\
      
      &  \multirow{1}{*}{ 3} &  \multirow{1}{*}{86\%}  & \multirow{1}{*}{0.78}  & \multirow{1}{*}{1.00}   & \multirow{1}{*}{0.88} \\
      \hline
              
\multirow{3}{*}{\makecell{Autoencoder~\cite{gunter2021compromised}}} 
        &  \multirow{1}{*}{ 1} &  \multirow{1}{*}{61\%}  & \multirow{1}{*}{0.56}  & \multirow{1}{*}{1.00}   & \multirow{1}{*}{0.72} \\
        
      &  \multirow{1}{*}{ 2} &  \multirow{1}{*}{61\%}  & \multirow{1}{*}{0.56}  & \multirow{1}{*}{1.00}   & \multirow{1}{*}{0.72} \\
      
      &  \multirow{1}{*}{ 3} &  \multirow{1}{*}{61\%}  & \multirow{1}{*}{0.56}  & \multirow{1}{*}{1.00}   & \multirow{1}{*}{0.72} \\
      
      \hline
      \multirow{3}{*}{\makecell{GAN-LSTM~\cite{li2022detecting}}} 
        &  \multirow{1}{*}{ 1} &  \multirow{1}{*}{66\%}  & \multirow{1}{*}{0.62}  & \multirow{1}{*}{0.81}   & \multirow{1}{*}{0.70} \\
        
      &  \multirow{1}{*}{ 2} &  \multirow{1}{*}{75\%}  & \multirow{1}{*}{0.67}  & \multirow{1}{*}{1.00}   & \multirow{1}{*}{0.79} \\
      
      &  \multirow{1}{*}{ 3} &  \multirow{1}{*}{66\%}  & \multirow{1}{*}{0.62}  & \multirow{1}{*}{0.80}   & \multirow{1}{*}{0.71} \\

          \bottomrule
          
    \end{tabular}
    \end{center}
    \label{tab:models}
    \vskip-10pt
\end{table}

\section{Conclusions}\label{section6}

In this article, we have considered three types of candidate cyberattacks on AVs with low levels of automation, i.e., ACC vehicles. We study the impacts of these attacks on both microscopic and macroscopic traffic flow dynamics. Motivated by these impacts, we then develop a machine learning based approach, i.e., a GAN-based model, for real-time detection of potential attacks on ACC vehicles. The proposed model is holistically evaluated considering all three types of attacks introduced. Numerical results on a set of candidate attacks selected for illustration show that the proposed model can effectively detect the vast majority of attacks with only a few misclassifications. \edit{The model developed is also observed to outperform some other existing ones in detecting abnormal ACC vehicle trajectory.}

This study offers important insights into understanding the potential impacts that a small number of compromised AVs (or ACC vehicles) could have on individual vehicles as well as on the bulk traffic. In addition, the proposed detection model provides a new way of identifying compromised vehicles in dynamic traffic flow, paving the way for building a safer, more resilient, and more sustainable transportation system as vehicles with autonomous driving capabilities become increasingly prevalent. While the methodology developed is promising, there is room for {further} improvement. Some directions for future work beyond \edit{the present study} are listed below:

\begin{itemize}

\item {While we have considered cyberattacks commonly seen in the literature, it is expected that new forms of attacks may emerge as advanced vehicle sensing and automation technologies are rapidly developing.} Hence, it would be interesting to look into other potential attacks as AV technologies advance. 

\item The detection model developed is limited in the length of time-series data it can use for detection. For instance, the model trained on 2 seconds of input data can only handle new observations \edit{with a length of 2 seconds.} \edit{Future work may seek to improve the proposed model towards a higher degree of generalizability for effectively handling different lengths of vehicle trajectory observed.}

\item \edit{Based on the detection of potential attacks}, it would be useful to develop robust and resilient control strategies for intelligent vehicles to ensure safe and stable traffic flow in the presence of attacks. 

\end{itemize}

\bibliographystyle{unsrt}
\bibliography{refs}
\end{document}